\documentclass[journal=nalefd,manuscript=letter,layout=onecolumn]{achemso}
\setkeys{acs}{maxauthors = 0}       
\setkeys{acs}{articletitle = false} 

\usepackage{chemformula} 
\usepackage[T1]{fontenc} 
\usepackage{amssymb}
\usepackage{amsmath}

\newcommand*\VEC[1]{\boldsymbol{#1}}

\author{Guan-Hao Peng}
\author{Ping-Yuan Lo}
\author{Wei-Hua Li}
\author{Yan-Chen Huang}
\author{Yan-Hong Chen}
\affiliation[National Chiao Tung University]
{Department of Electrophysics, National Chiao Tung University, Hsinchu 300, Taiwan, Republic of China}
\author{Chi-Hsuan Lee}
\author{Chih-Kai Yang}
\affiliation[National Chengchi University]
{Graduate Institute of Applied Physics, National Chengchi University, Taipei 11605, Taiwan, Republic of China}
\author{Shun-Jen Cheng}
\affiliation[National Chiao Tung University]
{Department of Electrophysics, National Chiao Tung University, Hsinchu 300, Taiwan, Republic of China}
\email{sjcheng@mail.nctu.edu.tw}

\title{Distinctive signatures of the spin- and momentum-forbidden dark exciton states
       in the photo-luminescence of strained WSe$_2$ monolayers under thermalization}


\begin{document}


\begin{tocentry}

\includegraphics{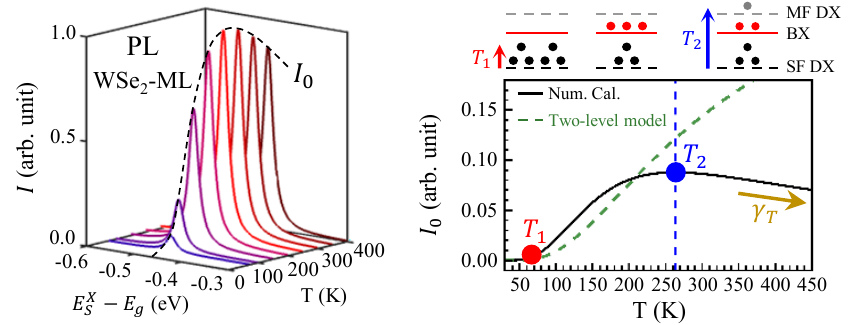}

\end{tocentry}

\begin{abstract}
 With both spin and valley degrees of freedom, the low-lying excitonic spectra of photo-excited transition-metal dichalcogenide monolayers (TMDC-MLs) are featured by rich fine structures, comprising the intra-valley bright exciton states as well as various intra- and inter-valley dark ones. The latter states can be classified as those of the spin- and momentum-forbidden dark excitons according to the violated optical selection rules.  Because of their optical invisibility, these two types of the dark states are in principle hardly observed and even distinguished in conventional spectroscopies although their impacts on the optical and dynamical properties of TMDC-MLs have been well noticed. In this Letter, we present a theoretical and computational investigation of the exciton fine structures and the temperature-dependent photo-luminescence spectra of strained tungsten diselenide monolayers (WSe$_2$-MLs) where the intra-valley spin-forbidden dark exciton lies in the lowest exciton states and other momentum-forbidden states are in the higher energies that are tunable by external stress. The numerical computations are carried out by solving the Bethe-Salpeter equation for an exciton in a WSe$_2$-ML under the stress-control in the tight-binding scheme established from the first principle computation in the density functional theory. According to the numerical computation and supportive model analysis, we reveal the distinctive signatures of the spin- and momentum-forbidden exciton states of strained WSe$_2$-MLs in the temperature-dependent photo-luminescences and present the guiding principle to infer the relative energetic locations of the two types of dark excitons.

\end{abstract}

\section{INTRODUCTION}

Atomically thin transition-metal dichalcogenide monolayers (TMDC-MLs) have recently drawn broad attention because of the extraordinary electronic and photonic properties \cite{mak2010atomically, mak2016photonics,wang2018colloquium,mueller2018exciton}. In contrast to the bulk counterparts, TMDC-MLs possess direct band gaps opened at two distinctive K- and K$'$-valleys in the reciprocal space \cite{mak2010atomically}, which are revealed as a new and optically accessible spin-like degree of freedom being the base of the emergent application of valleytronics \cite{sallen2012robust,zeng2012valley}. Moreover, because of the weakly screened Coulomb interaction in the low-dimensionality, the binding energy of an exciton in TMDC-MLs under photo-excitation is so high as hundreds of meV, 1-2 order of the magnitude higher than that of bulk semiconductors, \cite{he2014tightly,chernikov2014exciton} leading to the large exciton dipole moment and strong light-matter interaction \cite{lundt2016room,liu2017control}. The pronounced excitonic nature of photo-excited TMDC-MLs gives rise to fascinating physical phenomena, such as the room-temperature formation of exciton-polariton, \cite{lundt2016room} fast energy transfer rate, \cite{kozawa2016evidence} and high efficiency of light harvest, \cite{bernardi2013extraordinary} supportive for the prospective optoelectronic and photonic applications based on the emergent 2D materials.

However, the pronounced excitonic nature of TMDC-MLs seems not always to sustain efficient quantum yields and high radiative rates, both of which are required by advanced photonic and optoelectronic applications \cite{mak2016photonics,mueller2018exciton,baugher2014optoelectronic,wang2017efficient}. Without proper chemical treatment \cite{amani2015near,han2016photoluminescence}, the photo-luminescence and electro-luminescence yields of TMDC-MLs are typically so low as $\lesssim 1\%$ \cite{mak2010atomically,withers2015light,wang2017efficient,sundaram2013electroluminescence,malic2018dark}. The measured radiative times of TMDC-MLs are thus shown in the time scale $t> 1$ps, not as short as inferred from the intrinsic exciton dipole ($\sim 10^{-1}$ps) \cite{robert2016exciton,palummo2015exciton}. Both features are related to the effects from the optically inactive exciton states \cite{selig2018dark,chow2017phonon,schaibley2015population,zhang2018optically} and non-radiative channels \cite{robert2016exciton, shi2013exciton,palummo2015exciton}. Usually, the latter is set up via the extrinsic factors such as exciton-exciton collisions involving defects \cite{wang2014ultrafast} and lattices \cite{sun2014observation,shree2018observation}, while the former is from the intrinsic aspect of excitonic systems and becomes especially important as their energies are nearby those of the bright exciton (BX) states \cite{zhang2015experimental}.The darkness of the optically inactive exciton states is due to the violation of the spin and/or momentum selection rules that prohibit the optical creation or annihilation of such exciton states.  According to the violated rules, two classes of dark exciton (DX) states are specified, referred to as the spin-forbidden (SF) and momentum-forbidden (MF) dark excitons \cite{malic2018dark}.

Few years ago, Zhang \textit{et al.} \cite{zhang2015experimental} revealed the signatures about the SF DX states being the exciton ground states of the WSe$_2$-MLs in the measured temperature-dependent photo-luminescences (PLs) that are integrated over a long time period (up to few nanoseconds) in which the excitons are under thermalization. With the dark exciton ground states, the PL intensity of the measured WSe$_2$-ML is quenched at low-temperature, and shown increasing with increasing the temperature. Using the two-level model for simulating the thermal population of the BX states and the resulting temperature-dependent PLs, the existence of the SF DX ground states below the BX states by $\sim 30$meV is inferred. To directly measure the energies of the low-lying SF DX states in optical spectroscopies, one can apply magnetic fields to the tungsten-based TMDC-MLs and alter the carrier spins in the SF DXs to active the dipole moments, as reported by refs~\citenum{zhang2017magnetic,molas2017brightening,robert2017fine}. The measured energies of the SF DX states of WSe$_2$-MLs using the magneto-PL spectroscopy technique are shown in similar energy ($40-50$meV) to that inferred from the time-integrated PL measurement. The optical activation of the DX states by magnetic fields accounts for that the DX ground states of the WSe$_{2}$-ML belong to the SF DXs. Alternatively, by means of the near-field coupling to surface plasmon polaritons the SF DX states in the tungsten-based TMDC-MLs in the proximity of plasmonic materials can be activated and turn out to be measurable in optical spectroscopy \cite{zhou2017probing,park2018radiative}.

Besides the SF DX states, much attention is currently being paid to the MF DX states of TMDC-MLs, which are known deeply involved in the exciton dynamical processes via the efficient carrier-phonon interactions and impact the PL yields \cite{zhang2015experimental, selig2018dark}, the radiative lifetimes \cite{jakubczyk2016radiatively} and the spectral line width \cite{chow2017phonon,selig2016excitonic,christiansen2017phonon}. Those MF DX states, comprising the intra-valley exciton states resident out of the light cone and various inter-valley DX states, cannot be activated optically by applying magnetic field and are hardly distinguished from those of the BX and SF DX in the optical spectroscopy. Thus, alternative approaches, including the pump-probe spectroscopy \cite{berghauser2018mapping}, the cryogenic PL spectroscopy enabling the detection of the fine phonon replicas \cite{lindlau2017identifying} and strain-dependent PL spectroscopy \cite{hsu2017evidence}, have been employed to detect those MF DXs in TMDC-MLs. Nevertheless, the simultaneous diagnoses of the existences of both SF and MF DXs and further distinguishability of their distinctive energy locations yet remain as a desired but challenging task \cite{lindlau2017identifying}.
Theoretically, Malic {\it et al.} presented the computational investigation of the BX and DX band structures and the PL quantum yields of TMDC-MLs by solving the Wannier equation \cite{malic2018dark}, which well accounts for the observed low-temperature quenched behavior of the PL intensities of a WSe$_2$-ML \cite{arora2015excitonic}. Taking the similar approach where the {\it e-h} direct (exchange) interaction is considered (neglected or empirically parametrized), Selig {\it et al.} presented a microscopic study of the time-resolved formation and thermalization of the bright and dark excitons in TMDC-MLs and highlighted the role of the inter-valley dark exciton on the resultant temperature-dependent PL's \cite{selig2018dark}.

In this work, we present a theoretical and computational investigation of the exciton fine structures, composed of the states of the intra-valley BX as well as various SF and MF DXs, of strained WSe$_2$-MLs, by solving the Bethe-Salpeter equation (BSE), with the full consideration of the both electron-hole direct and exchange interactions, on the base of the tight-binding scheme established from the density functional theory (DFT). Based on the calculated exciton band structures, the temperature-dependent time-integrated PL from a WSe$_2$-ML that is determined by the thermal population on the BX states under the influence of the nearby SF and MF DX states, is investigated.  As the main result, the study reveals that, besides the low-temperature signatures about the lowest SF DX states, the time-integrated PLs in the high-temperature regime disclose also the signatures about the high lying MF DX states, which comprise the intra-valley and inter-valley MF DX ones. Under the influence of the high-lying MF DX states nearby the BX ones, the main PL intensity turns to descend with increasing the temperature in the room-temperature regime, where from the descending rate and turning temperature one can infer the existence and the energetic locations (above or below the BX states) of the MF DX bands. In addition to the computational studies, we conduct the analysis for the BX thermal population in a three-level model and explicitly derive the conditions for the emergence of the uncommonly known MF DX signatures, which are the multi-fold degeneracy and the proper energy order of the bright and dark excitons.  For further manifestation, we finally introduce a bi-axial strain in a WSe$_2$-ML as a tuning parameter to tailor the energy orders of the BX, SF and MF DX exciton bands and confirm the guiding principle for identifying and distinguishing the signatures of the two types of DXs.

\section{Results and discussion}

\subsection{Theoretical methodology}

In this work, we calculate the exciton spectra of TMDC-MLs by solving the BSE based on the DFT-calculated electronic band structures. Because the high numerical demand by both DFT and BSE computations, such an integrated DFT-BSE computation remains as a numerically large-scaled task and hard to implement. Thus, in order to reduce the numerical cost we partition the entire DFT-BSE computation procedure into different computational stages, each of which can be handled with feasible numerical effort, and at the end combined together in a physically reasonable manner. The partitioned computational procedure consists of the DFT computation, wannierization of the wave functions \cite{mostofi2008wannier90,mostofi2014an}, construction of the DFT-based tight binding Hamiltonian matrix \cite{kosmider2013large,scharf2016excitonic,lado2016landau}, and solving the BSE in the tight binding scheme \cite{ridolfi2018excitonic,trolle2014theory,berkelbach2015bright,scharf2016excitonic}.

\subsubsection{Band structure calculation using DFT}
First, we perform the DFT-computation for the quasi-particle band structures of the TMDC-MLs under study using the \textbf{\footnotesize{VASP}} package \cite{kresse1996efficient}. In the DFT, the quasi-particle band structure describing an electron in a crystalline material is determined by solving the Kohn-Sham equation
\begin{align}\label{eqn:Hdft}
  H^{KS} | \psi_{n\VEC{k}} \rangle = \epsilon_{n\VEC{k}} | \psi_{n\VEC{k}} \rangle \, ,
\end{align}
where $H^{KS}$ represents the Kohn-Sham Hamiltonian, $\psi_{n\VEC{k}}$ is the eigen state of Bloch function labeled by the band index $n$ and wave vector $\VEC{k}$, and  $\epsilon_{n\VEC{k}}$ is the energy of the eigen state \cite{kohn1965self}.
In this work, the DFT computation is carried out mainly using the hybrid exchange-correlation functional in the Heyd-Scuseria-Ernzerhof (HSE) version. \cite{heyd2003hybrid,heyd2004efficient,krukau2006influence}
The commonly used functional in the local density approximation (LDA) \cite{ceperley1980ground}, and the hybrid one in the Perdew-Burke-Ernzerhof (PBE) \cite{perdew1996generalized} version are employed as well, whenever the comparison with the previous DFT studies using those functionals \cite{kormanyos2015k} is needed.
The electron-ion interaction is represented by the projector-augmented wave (PAW) potential. The spin-orbital interactions (SOIs) are considered in the all computations throughout this work.
For more detailed technical information about the DFT computation, one might refer the Supporting Information (SI).
 Figure~\ref{fgr:QuasiParticleBandStructure} shows the calculated quasi-particle band structures of a WSe$_2$-ML using the computational approach.

\subsubsection{Wannierization}
Next, we transform the set of DFT-calculated electron wave functions, $\{ | \psi_{n\VEC{k}} \rangle \}$, into that of maximally localized Wannier functions (MLWFs), $\{ | \mathcal{W}_{i\VEC{R}} \rangle \}$, by using the well-established \textbf{\footnotesize{WANNIER90}} package \cite{mostofi2008wannier90,mostofi2014an}, where $i=(I,\alpha)$ is the composite index composed of the sites of the atoms positioned at $\VEC{\tau}_I$ in the primitive cell and  the atomic orbitals ($\alpha=p_x,p_y,p_z, d_{x^2-y^2}, d_{xy},...$) and   $\VEC{R}$ is the Bravais lattice vector.  In terms of the transformed MLWFs, the $\VEC{k}$-dependent Bloch-sum functions are defined by
\begin{equation}
| \phi_{i\VEC{k}} \rangle = \frac{1}{\sqrt{N}}\sum_{\VEC{R}}e^{i\VEC{k}\cdot\VEC{R}} | \mathcal{W}_{i\VEC{R}} \rangle \, ,
\end{equation}
which are used as the basis of linear combination of atomic orbitals (LCAO) to reformulate the DFT Hamiltonian as a matrix whose elements are given by
\begin{align}\label{eqn:TBHamiltonianMatrixElement}
H_{ij}^{TB}(\VEC{k})=\sum_{\VEC{R}} {e^{i\VEC{k}\cdot\VEC{R}}} t_{ij}^w (\VEC{R}) \, ,
\end{align}
where $t_{ij}^w (\VEC{R}) \equiv \langle \mathcal{W}_{i\VEC{0}} | H^{KS} | \mathcal{W}_{j\VEC{R}} \rangle $. The explicit form of eq~\ref{eqn:TBHamiltonianMatrixElement} is compatible with that of empirical tight-binding Hamiltonian in LCAO presentation \cite{slater1954simplified}, and allows us to relate the matrix elements $t_{ij}^w$ to the hopping and on-site parameters used in the latter \cite{kosmider2013large,scharf2016excitonic,lado2016landau}.

Once the non-diagonalized matrix of the DFT Hamiltonian is given, one can recalculate the quasi-particle band structure of the same material by carrying out the matrix diagonalization, and express the solved eigen states, in terms of the MLWFs, as
\begin{align}\label{eqn:QuasiParticleWavefunction}
  | \psi_{n\VEC{k}} \rangle =\sum_{i}^{N_{b}}C_{i}^{(n)}(\VEC{k}) | \phi_{i\VEC{k}} \rangle
  =\frac{1}{\sqrt{N}}\sum_{i}^{N_{b}}\sum_{\VEC{R}}^{N}C_{i}^{(n)}(\VEC{k})e^{i\VEC{k}\cdot\VEC{R}} | \mathcal{W}_{i\VEC{R}} \rangle \, ,
\end{align}
where $N_{b}=2\sum_{I=1}^M N_{a,I}$ is the total number of the electronic bands under the study, $N_{a,I}$ ($M$) is the number of chosen atomic orbitals for the $I-$th atom (the total number of the atoms) in the primitive cell, the factor 2 comes from the electron spin, and $N$ is the total number of primitive cells in the system. For a WSe$_2$-ML, a primitive cell contains one W-atom (labelled by $I=1$) and two Se-ones ($I=2,3$). Considering the five $d$-orbitals (the three $p$-orbitals) for the W-atom (Se-atom), we have $N_{a,1}=5$ and $N_{a,2}=N_{a,3}=3 $, and the total number of the orbitals in a primitive cell is $N_{b}=2\times (5+3+3)=22$.  In practical numerical implementation, the total number of the considered primitive cells that determine the total basis number is limited by the numerical resource and cannot be taken to be ideally infinite or so many as $N$. Thus, we truncate the number of the primitive cells considered in the computation by a finite number $N_R$, i.e. $\sum_{\VEC{R}}^{N} \rightarrow \sum_{\VEC{R}}^{N_R}$, between which the all electron hoppings are fully considered in eq~\ref{eqn:QuasiParticleWavefunction}, and take the number of $N_R$ as great as possible so that the calculated band structures quantitatively well converge to those directly calculated by the \textbf{\footnotesize{VASP}} package. In the sense, the sufficiently large matrix of Hamiltonian yielded from the wannierization transformation is equivalent to the original DFT-Hamiltonian. In this work, we take $N_R = 105 $, which yields the total number of the used MLWF basis $N_R \times N_b = 105 \times 22 = 2310$.

\subsubsection{Exciton energy spectrum: the Bethe-Salpeter equation}
Based on the calculated electronic band structures, the wave function of an exciton with the center-of-mass wave vector $\VEC{k} _{ex}$ can be written as
\begin{equation}
| \Psi _{S} \left( \VEC{k} _{ex} \right) \rangle = \frac{1}{\sqrt{\mathcal{A}}} \sum _{v c \VEC{k}} A _{S , \VEC{k} _{ex}} ^{v c} \!\! \left( \VEC{k} \right) \hat{c} _{c \VEC{k} + \VEC{k} _{ex}} ^{\dagger} \hat{h} _{v -\VEC{k}} ^{\dagger} | GS \rangle \, ,
\end{equation}
a linear combination of the configurations of the electron-hole ({\it e-h} ) pairs, $\hat{c} _{c \VEC{k} + \VEC{k} _{ex}} ^{\dagger} \hat{h} _{v -\VEC{k}} ^{\dagger} | GS \rangle $,
where the particle operator $\hat{c}_{c\VEC{k}}^{\dagger}$ ($\hat{h}_{v-\VEC{k}}^{\dagger}$) acting on the vacuum state $| GS \rangle$ is defined to create the electron (hole) of the wave vector $\VEC{k}$ ($-\VEC{k}$) in the conduction band $c$ (corresponding to the missing state at $\VEC{k}$ in the filled valence band $v$),
$S$ is the index of the exciton band, $A _{S, \VEC{k} _{ex}} ^{v c} \!\! \left( \VEC{k} \right)$ is the amplitude of the {\it e-h} configuration in the exciton wave function, and $\mathcal{A}$ is the area of the TMDC-ML material. Note that the spin states of the conduction and valence bands and those of the exciton states as well are implicitly absorbed in their band indices, $c$, $v$, and $S$, respectively.
As we are concerned with only the low-lying exciton states in this work, we designate $S=\rm{SA, SF}$ for the lowest two spin-resolved exciton bands, i.e. the spin-allowed (SA) and spin-forbidden (SF) ones, respectively.
The BSE for the determination of the exciton state reads\cite{sham1966many,hanke1974dielectric,hanke1980many,qiu2015nonanalyticity,wu2015exciton,zhang2017magnetic,echeverry2016splitting,deilmann2017dark,jiang2018tunability}
\begin{align}\label{eqn:BSE}
  \left[ \epsilon _{c \VEC{k} + \VEC{k} _{ex}} - \epsilon _{v \VEC{k}} \right] A _{S, \VEC{k} _{ex}} ^{v c} \!\! \left( \VEC{k} \right) + \sum _{v ^{\prime} c ^{\prime} \VEC{k} ^{\prime}} U \left( v c \VEC{k} , v ^{\prime} c ^{\prime} \VEC{k} ^{\prime} ; \VEC{k} _{ex} \right) A _{S ,\VEC{k} _{ex}} ^{v ^{\prime} c ^{\prime}} \!\! \left( \VEC{k} ^{\prime} \right) = E _{S}^{X} \left( \VEC{k} _{ex} \right) A _{S ,\VEC{k} _{ex}} ^{v c} \!\! \left( \VEC{k} \right)\, ,
\end{align}
where $E_{S}^{X}(\VEC{k}_{ex})$ is the energy of the exciton state, the first term is the total kinetic energy of the {\it e-h} pair composed of $\epsilon_{c\VEC{k}+\VEC{k}_{ex}}$ and $-\epsilon_{v\VEC{k}}$, and the second term is the kernel of {\it e-h} Coulomb interaction,
\begin{align}\label{eqn:ElectronHoleCoulombInteraction}
  U\left(vc\VEC{k},v'c'\VEC{k}';\VEC{k}_{ex}\right)
  =-V^{d}\left(vc\VEC{k},v'c'\VEC{k}';\VEC{k}_{ex}\right)
   +V^{x}\left(vc\VEC{k},v'c'\VEC{k}';\VEC{k}_{ex}\right),
\end{align}
composed of the screened {\it e-h} direct Coulomb interaction \cite{andersen2015dielectric,latini2015excitons,trolle2017model,hybertsen1986electron},
{\begin{equation}\label{eqn:DirectTermOfTheElectronHoleCoulombInteraction}
V ^{d} \left( v c \VEC{k} , v ^{\prime} c ^{\prime} \VEC{k} ^{\prime} ; \VEC{k} _{ex} \right) = \, \int d ^{3} \VEC{r} _{1} d ^{3} \VEC{r} _{2} \, \psi _{c \VEC{k} + \VEC{k} _{ex}} ^{*} \! \left( \VEC{r} _{1} \right) \psi _{v  \VEC{k}} \! \left( \VEC{r} _{2} \right) W \left( \VEC{r} _{1},\VEC{r} _{2} \right) \psi _{v ^{\prime}  \VEC{k} ^{\prime}} ^{*}  \! \left( \VEC{r} _{2} \right) \psi _{c ^{\prime} \VEC{k} ^{\prime} + \VEC{k} _{ex}}  \! \left( \VEC{r} _{1} \right)  \, ,
\end{equation}
and the {\it e-h} exchange one \cite{sham1966many,zhang2017magnetic,echeverry2016splitting,deilmann2017dark,jiang2018tunability,qiu2015nonanalyticity,guo2018exchange},

\begin{equation}\label{eqn:ExchTermOfTheElectronHoleCoulombInteraction}
V ^{x} \left( v c \VEC{k} , v ^{\prime} c ^{\prime} \VEC{k} ^{\prime} ; \VEC{k} _{ex} \right) = \, \int d ^{3} \VEC{r} _{1} d ^{3} \VEC{r} _{2} \, \psi _{c \VEC{k} + \VEC{k} _{ex}} ^{*} \! \left( \VEC{r} _{1} \right) \psi _{v \VEC{k}}  \! \left( \VEC{r} _{1} \right) V \left( \VEC{r} _{1} - \VEC{r} _{2} \right) \psi _{v ^{\prime} \VEC{k} ^{\prime}} ^{*} \! \left( \VEC{r} _{2} \right) \psi _{c ^{\prime} \VEC{k} ^{\prime} + \VEC{k} _{ex}}  \! \left( \VEC{r} _{2} \right) \, ,
\end{equation}
where $V \left( \VEC{r}_1 - \VEC{r}_2 \right)=\frac{e^{2}}{4\pi\varepsilon_{0}|\VEC{r}_{1}-\VEC{r}_{2}|}$ and $W \left( \VEC{r}_1, \VEC{r}_2 \right)$ are the bare and screened Coulomb potentials, respectively.
 Following the theory in refs~\citenum{andersen2015dielectric,latini2015excitons,trolle2017model,hybertsen1986electron}, the screened Coulomb potential is written as $\displaystyle W \left( \VEC{r}_{1} , \VEC{r}_{2} \right) = \int d ^{3} \VEC{r}' \varepsilon^{-1}\left( \VEC{r}_{1} , \VEC{r}' \right) V \left( \VEC{r}' - \VEC{r}_{2} \right)$ in terms of the inverse of the dielectric function $\varepsilon(\VEC{r}_{1},\VEC{r}_{2})$, which can be evaluated in the DFT by using the open source of ref~\citenum{database}.
 The technical details in the numerical calculations of eqs~\ref{eqn:DirectTermOfTheElectronHoleCoulombInteraction} and \ref{eqn:ExchTermOfTheElectronHoleCoulombInteraction} are given in SI. Remarkably, the short range part of eq~\ref{eqn:ExchTermOfTheElectronHoleCoulombInteraction} that essentially involves the quickly varying Wannier functions is hardly computed by using the common numerical integration method based on
the Newton-Cotes formulas, like the trapezoid rule, Simpson's rule, etc. To solve the problem, a semi-analytical method with the utilization of the cubic spline interpolation is developed in this work and used to calculate the integral of eq~\ref{eqn:ExchTermOfTheElectronHoleCoulombInteraction} at high accuracy (See SI for more details).

To solve the BSE of eq~\ref{eqn:BSE}, we discretize the exciton wave function on dense $k$-mesh points ($\sim 100\times 100$ in the first Brillouin zone) for the two lowest conduction ($c=c_1,c_2$) and the two highest valence bands ($v=v_1,v_2$) and express the BSE in the form of matrix-vector product. Then, the eigen energies (exciton energies) and eigen-vectors (exciton wave functions) of eq~\ref{eqn:BSE} are solved by carrying out the diagonalization for the matrix in the eigen value equation.
The validity of the employed methodology for solving the BSE is confirmed by the quantitative agreement between the BSE-calculated exciton bands and the low energy bands given by the effective exciton Hamiltonian explicitly derived in the dipole-dipole approximation \cite{yu2014dirac,qiu2015nonanalyticity,pikus1971exchange,ivchenko2005optical}, as presented in Figure~S4 of SI.

\subsubsection{Temperature-dependent photo-luminescences}

In the time-integrated PL over a long time up to few nanoseconds, excitons in a TMDC-ML at a specific temperature are considered to be at thermal equilibrium and populated in the excitonic states, including BX and various DX states, following Boltzmann statistics \cite{zhang2015experimental}. Thus, the intensity of the main PL peak is determined by the fraction of the BX states following the thermal statistics, which must possess the spin and momenta allowing for the optical transitions.
The PL spectrum from an exciton in a 2D material is calculated by using the Fermi's golden rule, as given by
\begin{align}\label{eqn:PLIntensity}
  I(\omega,T) \propto \sum_{S \in SA} \,\, \int\limits_{|\VEC{k}_{ex}|<{k_{c}}}d^{2}\VEC{k}_{ex} \, P(S,\VEC{k}_{ex}; T) |\langle 0 |\hat{P}_{\hat{e}}^{-}|\Psi_{S}(\VEC{k}_{ex})  \rangle |^2 \delta(E_{S}^{X}(\VEC{k}_{ex}) -\hbar \omega)\, ,
\end{align}
where $P(S,\VEC{k}_{ex}; T)$ is the thermal population of the exciton state $\Psi_{S}(\VEC{k}_{ex})$ at temperature $T$ and the operator $\hat{P}_{\hat{e}}^{-}\equiv \sum_{ij} p_{ij}^{\hat{e}}\hat{c}_{i} \hat{h}_{j}$ is defined to measure the dipole moment of the exciton state with respect to the $\hat{e}$-polarized light and in terms of the momentum matrix elements (MMEs), $p_{ij}^{\hat{e}} \equiv \hat{e}\cdot \langle \psi_{j}| \VEC{p} |\psi_{i} \rangle $ \cite{grosso2013solid,cohen2016fundamentals}. Equation \ref{eqn:PLIntensity} considers only the BX states that possess the carrier spins allowing for light emission ($S=SA$) and the momentum lying in the light cone defined by the cone boundary $k_c=\omega/c$,  where $c$ ($\omega$) is the speed (angular frequency) of the emitted light. \cite{palummo2015exciton}
In the dipole approximation, the MME with respect to a valence state and a conduction state of a TMDC-ML is given by \cite{pedersen2001optical,berkelbach2015bright}
\begin{align}\label{eqn:MME}
\langle \psi_{n'\VEC{k}} | \VEC{p} | \psi_{n\VEC{k}} \rangle
  \approx \frac{m_{e}}{\hbar}\sum_{I,J=1}^{M}\sum_{\alpha,\beta}C_{J\beta}^{(n')*}(\VEC{k})C_{I\alpha}^{(n)}(\VEC{k})
   \nabla_{\VEC{k}}H^{TB}_{J\beta,I\alpha}\left(\VEC{k}\right) \, ,
\end{align}
which is valid for generic TMDC-MLs where the inter-atom (intra-atom) dipole moments are pronounced (negligible) \cite{berkelbach2015bright}.
Assuming the Boltzmann statistics, the fractional population of a specific exciton state, $\Psi_{S}\left(\VEC{k}_{ex}\right)$, at the temperature $T$ is given by
\begin{align}\label{eqn:FractionalPopulationOfTheExcitonState}
P(S,\VEC{k}_{ex};T)=\frac{\exp[-E_{S}^{X}\left(\VEC{k}_{ex}\right)/k_{B}T]}{\displaystyle{\sum_{S'}\sum_{\VEC{k}'_{ex} \in \rm{BZ}} \exp[-E_{S'}^{X}\left(\VEC{k}'_{ex}\right)/k_{B}T]}} \, ,
\end{align}
where $\rm{BZ}$ indicates the first Brillouin zone (1st BZ) in the $\VEC{k}_{ex}$ space and $k_{B}$ is the Boltzmann constant.
As the BX states lie in a small area of light cone where $|\VEC{k}_{ex}|<k_c$ ($\sim 0.008 \rm{nm}^{-1}$), we might consider $\langle 0 |\hat{P}_{\hat{e}}^{-}|\Psi_{S}(\VEC{k}_{ex})  \rangle \approx \langle 0 |\hat{P}_{\hat{e}}^{-}|\Psi_{S}(\VEC{0})  \rangle$ and $\delta(E_{S}^{X}(\VEC{k}_{ex}) -\hbar \omega) \approx \delta(E_{S}^{X}(\VEC{0}) -\hbar \omega)$ in eq~\ref{eqn:PLIntensity}, and the intensity of the main PL peak $I_0 (T) \equiv I(\omega=E_{\rm{SA}}^{X} (\VEC{0})/\hbar; T)$ is approximated by
\begin{equation}\label{eqn:PLPeakIntensity}
  I_0(T) \propto  P_{\rm{BX}}(T) \equiv \sum_{\rm{SA}} \int\limits_{|\VEC{k}_{ex}|<{k_{c}}}d^{2}\VEC{k}_{ex} P\left(\rm{SA},\VEC{k}_{ex};T\right) \, ,
\end{equation}
which is simply proportional to the total population of the BX states in the light cone.

\subsection{Quasi-particle band structures}

Figure~\ref{fgr:QuasiParticleBandStructure}a presents the full-zone quasi-particle band structure of a WSe$_2$-ML calculated by using the \textbf{\footnotesize{VASP}} package in the HSE06-DFT, where the red (blue) lines indicate the spin-up (spin-down) states and the band indices $c=c_{1}, c_2,...$ ($v=v_{1},v_2,...$) are sorted by the band energies from low to high (high to low).
Same as other TMDC-MLs, the basic feature of the band structure of a WSe$_2$-ML comprises the direct band gaps opened at the $K$ and $K'$ valleys, and a giant spin-splitting in the valence band valley, $\sim$ 618.5 meV (a smaller spin-splitting in the conduction band valley, $\sim$ 3.4 meV), induced by the intrinsic SOI. The large energy difference between the valence- and conduction-SOI splittings originates from the different atomic natures of the microscopic Bloch functions, which are the $d$-like and $p$-like atomic orbitals governing the valence and the conduction states, respectively \cite{kosmider2013large,roldan2014momentum}. The giant spin-splitting in the valence band leads to two classes of exciton states that are spectrally well apart, i.e. the A-exciton and B-exciton in the low- and high-energy regime, respectively. Since our main interest is in the low-lying exciton states, here one might focus on the A-exciton relevant bands, i.e. the sole topmost non-degenerate valence band ($v_{1}$) and the lowest spin-split conduction bands ($c_{1}, c_{2}$).
Because of the spin-valley locking effect combined with the giant spin splittings in the valence bands \cite{xiao2012coupled}, the spin of the valence hole in a A exciton is fixed and firmly determined by the occupied valleys. Thus, the spin-related fine structure of the A-exciton states is essentially related to the spin structure of the conduction band. Figure~\ref{fgr:QuasiParticleBandStructure}b shows the zoomed-in view of the spin-resolved conduction and valence bands, where the SOI-induced splitting in the conduction band, $\Delta\epsilon_{c_{2},c_{1}}^{KK}\equiv{\epsilon_{c_{2}\VEC{K}}-\epsilon_{c_{1}\VEC{K}}}=3.4$meV, is clearly observed.
The calculated conduction splitting of a WSe$_{2}$-ML is consistent with those calculated in ref~\citenum{echeverry2016splitting} using the same HSE-DFT method. In this work, the adoption of the HSE functional in the DFT computation for WSe$_2$-ML is according to the studies of ref~\citenum{echeverry2016splitting}, which show that, as compared with other exchange-correlation functionals LDA and PBE, the HSE-DFT-calculated conduction splitting at the scale of some meV is in relatively better agreement with those from other sophisticated DFT-GW calculations \cite{rasmussen2015computational,kormanyos2015k,
kormanyos2014spin,echeverry2016splitting,danovich2016auger,wang2015spin,kosmider2013large,
ramasubramaniam2012large,druppel2018electronic,deilmann2017dark,zhang2017magnetic,eickholt2018spin,
kang2013band} as summarized in Figure S8a of SI. The use of the HSE-DFT method is supported also by the excellent prediction of the conduction splitting of another tungsten based TMDC-ML, WS$_2$-ML, well matching the experimental values measured by angle-resolved photoemission spectroscopy (ARPES) \cite{eickholt2018spin}. The statistics of the calculated and experimental values of the conduction splitting of WS$_2$-ML reported in literature is presented in Figure~S8b of SI.
Using the exchange-correlation functional in the LDA, the calculated quasi-particle band structure of a WSe$_{2}$-ML shows the bigger conduction splitting $\Delta\epsilon_{c_{2},c_{1}}^{KK}\sim 38$meV and the lower conduction $Q$-valleys as presented in Figure S1a of SI. One also notes that the calculated quasi-particle band structure of a WSe$_{2}$-ML using the PBE functionals is quantitatively very similar to the LDA-calculated one. 

Concerned with the optical selection rule, only the transition involving the conduction and valence states of the same electron spin is optically active and the other one involving the opposite spin is optically inactive. Note that the upper conduction band-edged state owns the same spin as that of the topmost valence state. Therefore, in the simple scenario at non-interacting level the lowest exciton states of WSe$_2$-ML are inferred to be the optically inactive SF DX states, while the next high exciton level could be that of the BX ones.

Figure~\ref{fgr:QuasiParticleBandStructure}c and d show the energy contours of the lowest conduction band $c_{1}$ and the highest valence band $v_{1}$ of a WSe$_2$-ML in the 1st Brillouin zone, respectively. In addition to the typical $K$ and $K'$ valleys, one notes the emergence of the additional six conduction-band local minima  that are located near to the midpoints between $\Gamma$- and $K(K')$-points, and  named by $Q_{i=1,2,3}$    ($Q_{i=1,2,3}^{'}$), respectively. The calculated energy difference between the conduction $K/K'$ and $Q_i/Q_i'$ valleys is given by $\Delta \epsilon_{c_{1},c_{2}}^{QK}\equiv{\epsilon_{c_{1}\VEC{Q}}-\epsilon_{c_{2}\VEC{K}}}=77.7$meV, falling in the similar energy scale as reported in previous works \cite{palummo2015exciton,rasmussen2015computational,le2015spin,amin2014strain,yun2012thickness,sahin2013anomalous}.

\subsection{Exciton band structures}

Figure~\ref{fgr:ExcitonBandStructure} (Figure~\ref{fgr:ExchangeExcitonBandStructure}) shows the calculated band structure of a single exciton in a WSe$_2$-ML by solving the BSE based on the quasi-particle band structure of Figure~\ref{fgr:QuasiParticleBandStructure} and with the neglect (consideration) of the {\it e-h} exchange interaction.
First, let us begin with Figure~\ref{fgr:ExcitonBandStructure} to recognize the main characteristics of the exciton band structure of a WSe$_2$-ML. 
 Offset by the non-interacting energy gap $E_{g}$, the energy values of Figure~\ref{fgr:ExcitonBandStructure} are equal to the additive inverse of the exciton binding energy.
In Figure~\ref{fgr:ExcitonBandStructure}, one notes that the exciton band structure of a WSe$_2$-ML exhibits the pronounced multi-valley feature. First, the intra-valley exciton, where the electron and hole are resident in the same $K$ or $K'$ valley, possesses the exciton wave vector $\VEC{k}_{ex}\approx \VEC{0}$, mapping to the $\Gamma_{X}$ point as depicted in Figure~\ref{fgr:ExcitonBandStructure}a, where for exciton the subscript $X$ is designated to the subscript of the symbols representing high-symmetry points in the reciprocal $\VEC{k}_{ex}$ space.
The inter-valley $KK'$ ($K'K$) exciton where the electron and hole are resident in the opposite $K$ and $K'$ valleys ($K'$ and $K$ valleys), respectively, possesses the large wave vector mapped to the $K_{X}$ or $K_{X}'$ corners of the first Brillouin zone, as indicated in the insets of Figure~\ref{fgr:ExcitonBandStructure}a and b. Because of the spin splitting in the conduction band, the intra-valley ($KK$ and $K'K'$) exciton bands are split into two bands, which are the lower SF DX band and the upper SA one, respectively.
Likewise, the inter-valley ($KK'$ and $K'K$ ) exciton bands are split into the SF and SA exciton bands but in the reversed energy order. Note that only a small fraction of the intra-valley SA exciton states that lie in the reciprocal area of the light cone are optically active and belong to the BX states.

The emergence of the conduction $Q$ valleys in the WSe$_2$-ML leads to the low-lying inter-valley exciton where the hole is in the $K$ or $K'$ valley and the electron is in one of the six $Q_{i}/Q_{i}'$ valleys. Thus, in total twelve inter-valley $KQ$ (i.e. $KQ_{i}$, $KQ_{i}'$, $K'Q_{i}$ and $K'Q_{i}'$, with $i=1,2,3$) exciton bands results. The three inter-valley $KQ_{i}$ ($K'Q_{i}'$) excitons possess the exciton wave vectors mapping to the three $Q'_{X,i}$ ($Q_{X,i}$) points and the other six inter-valley $KQ'_{i}$ and $K'Q_{i}$ excitons possess the exciton wave vectors mapping to the three $M_{X,i}$ points in the $\VEC{k}_{ex}$ space each of which is doubly degenerate, as illustrated in the insets of Figure~\ref{fgr:ExcitonBandStructure}a and b. In order to facilitate the later discussions,  the phrase "$KQ$ exciton" is used to stand for all the $Q_{X,i}$, $Q_{X,i}'$ and $M_{X,i}$ exciton bands afterward.

The excitonic effects in Figure~\ref{fgr:ExcitonBandStructure} are identified by the comparison with the  non-interacting transition energies, as indicated by the upward arrow lines connecting the valence and conduction valleys in Figure~\ref{fgr:QuasiParticleBandStructure}a and b.
First, the level energy of the intra-valley SF DX states is further lowered and the energy separation between the BX and the SF DX states is enlarged to be $\Delta_{\rm{SA},\rm{SF}}^{\Gamma \Gamma} \equiv E_{\rm{SA}}^{X}(\VEC{0}) - E_{\rm{SF}}^{X}(\VEC{0}) \approx 38$meV, which results from the larger exciton binding energy of the SF DX band that involves the $c_{1}$ conduction band because of the heavier effective mass of the $c_{1}$ band than that of the $c_{2}$ one at the $K/K'$ valleys.
The effective mass of the $n$-th band at $\VEC{k}_{0}$ is determined by using the least square method to fit the numerically calculated data at ten $\VEC{k}$ points around $\VEC{k}_{0}$ along the direction between $\Gamma-K$ and $\Gamma-M$ paths to the parabola $\epsilon_{n\VEC{k}}|_{\VEC{k}\approx \VEC{k}_{0}} = \epsilon_{n\VEC{k}_{0}} + \frac{\hbar^{2} |\VEC{k}-\VEC{k}_{0}|^2}{2 m_{n, \VEC{k}_{0}}}$, where the effective mass is implicitly defined under the assumption of band isotropy and parabolicity \cite{kormanyos2015k}.
Table~\ref{tbl:ElectronicMass} summarizes the effective masses of the quasi-particle band dispersions around the different valleys presented in Figure~\ref{fgr:QuasiParticleBandStructure}.
Throughout this work, the notation, $\Delta_{S',S}^{\VEC{k}_{0}'  \VEC{k}_{0} } \equiv E_{S'}^{X}(\VEC{k}_{ex}=\VEC{k}_{0}') - E_{S}^{X}(\VEC{k}_{ex}=\VEC{k}_{0})$, is used to indicate the energy difference between two exciton band-edged states, where $\VEC{k}_0 \in \{ \Gamma_X, Q_{X,i},Q'_{X,i}, M_{X,i}, K_X, K'_X \}$. The computed energy splitting, $\Delta_{\rm{SA},\rm{SF}}^{\Gamma \Gamma}=38$meV, between the intra-vaelly BX and SF DX exciton bands is surprisingly close to the measured value in ref~\citenum{zhang2017magnetic} in spite of the neglect of the {\it e-h} exchange interactions in the computation. This might result from the cancellation of the missing of the positive contribution of the {\it e-h} exchange interaction to $\Delta_{\rm{SA},\rm{SF}}^{\Gamma \Gamma}$ and the over-estimation of $\Delta\epsilon_{c_{2},c_{1}}^{KK}$ by the HSE functional (as compared with the sophisticated GW method \cite{echeverry2016splitting,zhang2017magnetic}), both of which are estimated to be in the energy scale of some meV but affect the energy splitting in the opposite manner.  As reported by ref~\citenum{echeverry2016splitting}, the GW correction makes lower the energy of the BX and decreases the $\Delta_{\rm{SA},\rm{SF}}^{\Gamma \Gamma}$ by $7$meV, while the inclusion of the {\it e-h} Coulomb interaction increases $\Delta_{\rm{SA},\rm{SF}}^{\Gamma \Gamma}$ by $6$meV at the similar energy scale \cite{zhang2017magnetic}.

Second, one notes that because of the enhanced exciton binding energy by the heavy mass of the $Q$-valleys, the difference between the band edges of the intra-valley $KK$ BX and the inter-valley $KQ$ DX is reduced to be only $\Delta_{\rm{SA},\rm{SA}}^{Q\Gamma} = \Delta_{\rm{SF},\rm{SA}}^{M\Gamma} \approx 16$meV, which is much smaller than the energy difference between the conduction $K$ and $Q$ valleys, $\Delta \epsilon_{c_1,c_2}^{QK}=77.7$meV. With the small energy difference from the BX states, the inter-valley MF DX states in WeS$_2$-MLs should be essentially involved in the exciton and optical properties \cite{chow2017phonon,selig2018dark}.

Next, let us further examine the exciton band structures of a WSe$_2$-ML under the influence of the {\it e-h} exchange interaction which are shown in Figure~\ref{fgr:ExchangeExcitonBandStructure} and Figure S4 in SI.
As the main effects on the exciton fine structures, the long range part of the {\it e-h} exchange interaction gives rise to the energy splitting of the bright exciton band ($\sim 2.28$meV at the light cone edge $k_{ex}=k_c$; See Figure S4 in SI), while the short range part is shown to further separate the levels of the bright exciton and the spin-forbidden dark one by 9meV and lifts the degeneracies of the intra- and inter-valley dark exciton states.
Consequently, with the {\it e-h} exchange interaction the calculated energy separation between the BX and the SF DX states is increased to  $\Delta_{\rm{SA},\rm{SF}}^{\Gamma \Gamma} \approx 47$meV, well falling in the energy range of the measured values $\Delta_{\rm{SA},\rm{SF}}^{\Gamma \Gamma, \rm{exp}} =40-50$meV \cite{zhang2017magnetic,molas2017brightening,zhou2017probing,wang2017plane,park2018radiative}.
The breaking of the degeneracy of the momentum-forbidden inter-valley DX bands arises from the short range part of the exchange interaction that makes the blue-shift of the spin-allowed exciton states but does not affect the spin-forbidden ones \cite{echeverry2016splitting}.
The band-edge levels of the calculated low-lying exciton bands of a WSe$_{2}$-ML without and with the consideration of the {\it e-h} exchange interaction are shown in the insets of Figure~\ref{fgr:TemperatureDependentPLIntensity}a and \ref{fgr:TemperatureDependentPLIntensity}b, respectively.
With the interaction-induced spin-splitting, it turns out that the all inter-valley MF-DX band edge levels (including the both $KQ$ and $KK'/K'K$ DX levels labeled by the grey dashed lines in Figure~\ref{fgr:TemperatureDependentPLIntensity}b) distribute over a wide energy range of $\sim 60$meV where the BX level (labeled by the red solid line)  lies in-between.
The identification of the lowest DX states of a WSe$_{2}$-ML in our theory is consistent with the experimental observations reported by refs~\citenum{zhang2017magnetic,molas2017brightening,robert2017fine}. The revealed exciton fine structure differs slightly from but is close to the recent experimental speculation according to the analysis of the cryogenic PL spectra of a WSe$_{2}$-ML measured by ref~\citenum{lindlau2017identifying}.

\subsection{Exciton thermalization and Temperature-dependent PL spectra}

Since we are mainly concerned with the thermally accessible low-lying states nearby the exciton valleys, with the neglect of the {\it e-h} exchange interaction one can characterize the local nearly parabolic exciton bands in terms of the effective masses and the band-edge energies of the exciton valleys, i.e. $E_{S}^{X}(\VEC{k}_{ex})|_{\VEC{k}_{ex} \sim \VEC{k}_{0}} \approx  E_{S}^{X}(\VEC{k}_{0}) + \frac{\hbar^2 |\VEC{k}_{ex}-\VEC{k}_{0}|^2}{2M_{S, \VEC{k}_{0}}^X}$, where $\VEC{k}_{0} $ are the wave vectors centered in the exciton valleys around ($\Gamma_X, Q_{X,i},Q_{X,i}^{'} ,M_{X,i}, K_{X},K_{X}^{'}$) in the reciprocal space.
Table~\ref{tbl:ExcitonicMass} summarizes the effective masses of the excitonic band extrema evaluated according to the band dispersions of Figure~\ref{fgr:ExcitonBandStructure}b.

In the approximation of parabolicity and isotropy, the population of BX under thermalization at the temperature $T$ according to eq~\ref{eqn:PLPeakIntensity} reads
  \begin{align}\label{eqn:FractionalPopulationOfTheBXStates}
  P_{BX}(T) = \frac{2 R_{k_{c}} M_{SA,\Gamma}^{X}}{\displaystyle{\sum_{S}\sum_{\VEC{k}_{0}}M_{S,\VEC{k}_{0}}^{X}\exp\{-[E_{S}^{X}(\VEC{k}_{0}) - E_{SA}^{X}(\VEC{0})]/k_{B}T\} }} \, ,
\end{align}
where the factor 2 in the numerator originates from the two-fold degeneracy ($KK$ and $K'K'$) of the intra-valley SA bands, $R_{k_{c}}= \left\{ 1 - \exp [-\hbar^{2} k_{c}^{2} / 2M_{SA,\Gamma_{X}}^{X} k_{B}T]\right\}$   determines the proportion of the BX states that lie in the light-cone area of the intra-valley SA exciton band characterized by the wave-vector boundary $k_c$.

Note that, in the presence of {\it e-h} exchange interaction, the bright exciton bands might not retain the band parabolicity, as known from refs \citenum{qiu2015nonanalyticity,wu2015exciton,yu2014dirac} and seen also in Figure~\ref{fgr:ExchangeExcitonBandStructure}. Under the situation, the thermal population of the exciton states should be evaluated using eqs \ref{eqn:FractionalPopulationOfTheExcitonState} and \ref{eqn:PLPeakIntensity}, instead of eq~\ref{eqn:FractionalPopulationOfTheBXStates}.
The intensities of the main PL peaks corresponding to Figure~\ref{fgr:TemperatureDependentPLIntensity}a and \ref{fgr:TemperatureDependentPLIntensity}b are presented in Figure~\ref{fgr:TemperatureDependentPLIntensity}c in comparison with each others.
Though the {\it e-h} exchange interaction does induce additional exciton fine structures as shown in the insets of Figure~\ref{fgr:TemperatureDependentPLIntensity}a and \ref{fgr:TemperatureDependentPLIntensity}b, the calculated PL intensities with and without the consideration of {\it e-h} exchange interaction exhibit the similar temperature dependences, shown clearly by the comparison of Figure~\ref{fgr:TemperatureDependentPLIntensity}a, \ref{fgr:TemperatureDependentPLIntensity}b, and \ref{fgr:TemperatureDependentPLIntensity}c. For facilitating the physical analysis, we thus consider only the the {\it e-h} direct Coulomb interaction and neglect the exchange one in the BSE for the main results presented hereafter.

According to eqs~\ref{eqn:PLIntensity} and \ref{eqn:FractionalPopulationOfTheBXStates}, Figure~\ref{fgr:TemperatureDependentPLIntensity}a shows the calculated PL spectra of a WSe$_2$-ML at the temperatures $T=75, 100, 150, ...400$K, where the $\delta$-function in eq~\ref{eqn:PLIntensity} is approximated by the Lorentzian function, $\delta(E-\hbar \omega) \approx \frac{1}{\pi} \frac{\gamma/2}{(E-\hbar\omega)^{2}+(\gamma/2)^{2}}$, with the phenomenological broadening $\gamma=25$meV. The intensity of the main PL peak as a function of the temperature is presented in Figure~\ref{fgr:TemperatureDependentPLIntensity}c. First, one sees that the PL intensity is quenched at low-temperature $T<50$K where the exciton population is mainly on the lowest intra-valley SF DX states that emit no light. Then, the PL intensity rises up rapidly as the temperature exceeds $T_1 \sim 66$K where excitons start to populate the BX states significantly. Such a $T$-dependence of PL intensity reflecting the thermal population of the BX states was commonly understood by means of simulation using the two-level model \cite{zhang2015experimental,hsu2017evidence}, as depicted in Figure~\ref{fgr:ThreeLevelModel}a, that comprises a lower DX level and an upper BX one with the energies (degeneracies) denoted by $E_{D_{1}}$ and $E_{B}$ ($g_{D_{1}}$ and $g_{B}$), respectively. In the two-level model, the rising population of the BX state with the increasing temperature around $T=T_1$ is described by
\begin{align}\label{twolevel}
  P_{BX}(T) = \left( \frac{g_{D_{1}}}{g_{B}} e^{\Delta_{1} \beta_{T}} + 1 \right)^{-1} \approx \frac{g_B}{g_{D_{1}}}e^{-\Delta_{1} \beta_T} \, ,
\end{align}
(see eqs S22 and S23 in SI) where $\Delta_{1} \equiv E_{B}-E_{D_{1}} >0$, $\beta_T \equiv 1/k_{B}T$, and $\frac{\Delta_{1}}{k_{B}T_{1}} \ll 1$ is considered. Fitted to the two-level model, the rising temperature can be related to the energy splitting between the DX and BX states via the equation, $T_1 \approx \frac{1}{5-\ln{(g_{D_{1}}/g_{B})}}\frac{1}{k_{B}}\Delta_{1}$, where the population of BX is set to be $P_{BX}(T=T_1) = e^{-5} \lesssim 1\%$ (See SI for the detailed derivation).

Following the two-level model, the BX population should keep increasing with increasing the temperature and eventually remain nearly constant at high-temperature where $T \gg \Delta_{1}/k_B$ \cite{zhang2015experimental}. However, one sees that the numerically simulated PL intensity in Figure~\ref{fgr:TemperatureDependentPLIntensity}c yet turns to descend at $T>T_{2} \sim 263$K.
The descending PL intensity at the high-temperature reflects that the BX states are somehow unusually depopulated with increasing temperature. This means that, with increasing the temperature,  some fraction of the BX states is increasingly transferred to some other exciton states, which should lie at higher energy and emit no light (i.e. dark states), and the out-transferred population is even more than that supplied from the lowest exciton states. As the probability of occupying a high energy state is always smaller than a low energy one, the high states can be populated more than the low one with increasing the temperature only if the number of the high energy states is greater than that of the low energy ones. This is exactly the case of the inter-valley $KQ$ exciton states in WSe$_2$-ML, which are optically inactive, highly (12-fold) degenerate, and energetically little higher than the BX according to our simulation.

To confirm the scenario, we carry out the analysis in a three-level model that comprises one BX-level and {\it two} DX-ones, with the energies (degeneracies) denoted by $E_{B}$, $E_{D_1}$, and $E_{D_2}$ ($g_{B}$, $g_{D_1}$, and $g_{D_2}$), respectively, as depicted in Figure~\ref{fgr:ThreeLevelModel}b. In the Boltzmann statistics, the population of the BX level in the three-level model is given by
\begin{align}\label{threelevel}
  P_{BX}(T)=\left( r_1 e^{\Delta_{1} \beta_T } + 1 + r_2 e^{-\Delta_{2} \beta_T} \right)^{-1} \, ,
\end{align}
where $r_{n}\equiv \frac{g_{D_n}}{g_{B}}$, $\Delta_1 \equiv E_{B} - E_{D_1}$ and $\Delta_2 \equiv E_{D_2} - E_{B}$.

Figure~\ref{fgr:ThreeLevelModel}c shows the fractional population of the BX level as a function of the temperature calculated from the three-level model for the cases with $\Delta_{1}=\Delta_{2}>0$ and $r_1=1$ and $r_2=1,3,6$, are chosen to simulate the excitonic structure of WSe$_2$-ML presented by Figure~\ref{fgr:ExcitonBandStructure}a. As we expect, the BX population does show the $T$-dependence as that of Figure~\ref{fgr:TemperatureDependentPLIntensity}c, featured by a low rising temperature $T_1$  and a high turning temperature $T_2$ at which the BX population starts to rise up rapidly and turns to descend slowly, respectively.
At a low $T$ ($\ll \Delta_{1}/k_B$), the BX population is derived as $P_{BX}(T)\approx \frac{g_{B}}{g_{D_1}}e^{-\Delta_1\beta_T}[1-\frac{g_{B}}{g_{D_1}}e^{-\Delta_1\beta_T}-\frac{g_{D_2}}{g_{D_1}}e^{-(\Delta_1+\Delta_2)\beta_T}]\approx \frac{g_{B}}{g_{D_1}}e^{-\Delta_1\beta_T}$ (see SI for details of the model analysis), approximately same as that given by the two-level model. Hence, the emergence of the low rising temperature $T_{1}$ evidences {\it only} the {\it lowest} DX states below the BX states and is irrelevant to other high-lying DX ones.

Besides, one realizes from the model analysis that the descending feature emerges especially as the second DX level ($D_2$) lies above the BX one.
This condition can be manifested by analyzing the change rate of the BX population with respect to temperature in the high-temperature limit, which can be shown to be
\begin{equation}
\gamma_T\equiv \frac{d P_{BX}}{dT}\propto (g_{D_1}(E_{B}-E_{D_1}) - g_{D_2}(E_{D_2}-E_{B}) )\, .
\end{equation}
Accordingly, the asymptotic change rate of the BX population is negative as $E_{D_2}> E_{B} + \frac{g_{D_1}}{g_{D_2}}(E_{B}-E_{D_1}) >E_B$.

Besides a significant descending rate, the descending feature is in practice easily observable if the turning temperature $T_2$ is low enough to fall into the room-temperature regime.
In the solvable three-level model, the turning temperature can be derived as
\begin{equation}
T_2 = \frac{\Delta_{1}}{k_B} \frac{\left(1+\frac{\Delta_{2}}{\Delta_{1}}\right)}{\ln (\frac{g_{D_2}}{g_{D_1}} \frac{\Delta_{2}}{\Delta_{1}}) } \, ,
\end{equation}
which is positive only if $\Delta_{2}>\frac{g_{D_1}}{g_{D_2}}\Delta_{1}$. From the formalism, it is shown that the turning temperature $T_2$ can be lowered by the high degeneracy of the high lying DX level, $g_{D_2}$, with respect to that of the lowest DX, $g_{D_1}$.
Figure~\ref{fgr:ThreeLevelModel}c shows the simulated BX population versus temperature by the three-level model with the different ratios of the BX and DX degeneracies. With the equal degeneracy of the three levels ($g_{D_1}=g_{D_2}=g_B$), the BX population always persist increasing with increasing the temperature and no $T_2$-characteristics is observed. Increasing the degeneracy of the high DX, say $g_{D_2}=3$,  the BX population keeps increasing  at the low-temperature $T \lesssim 1.5 \Delta_1/k_B$ and turn to descend as the temperature exceeds some critical temperature $T =T_2 \sim 1.5 \Delta_1/k_B$. Further raise the degeneracy of the high DX states to $g_{D_2}=6$, the descending feature at high-temperature becomes even more pronounced and characterized by the lowered $T_2$. The model simulation implies that the emergence of the observable $T_2$-signature requires the high degeneracy of the high lying DX states with the energy appropriately higher than or comparable to the BX ones.

To compare the theoretical prediction with experiments, in Figure~\ref{fgr:TemperatureDependentPLIntensity}c we place the measured data of temperature-dependent PL intensities duplicated from Figure~3a of ref~\citenum{zhang2015experimental}. The calculated temperature-dependent PL intensity of a WSe$_{2}$-ML is, both qualitatively and quantitatively, well consistent with the experimental observation. Remarkably, taking a closer look at the measured data at high temperatures $T>260$K, one does observe a descending tendency of the PL intensities with increasing the temperature, as predicted by our theory. In fact, the descending rate of the measured PL intensity with respect to the temperature is even faster than our theoretical prediction. This could be attributed to other extrinsic optically inactive states, e.g. surface or defective ones, or any non-radiative channels that likely exist in real samples but are beyond the scope of this work \cite{wang2014ultrafast,sun2014observation}.

In Figure S1 of SI, we show the calculated band structure, exciton band edge levels, and the temperature dependent PL spectrum of a WSe$_{2}$-ML based on the LDA-DFT calculated band structure. Using the LDA or PBE functionals in the DFT computation, the predicted exciton ground states of a WSe$_{2}$-ML remain as the intra-valley SF DX ones, but with the lower energy separated from the higher BX level, $\sim$ 80meV, which is slightly larger than the calculated value by ref \citenum{malic2018dark} because of  the effect of the {\it e-h} exchange interaction considered in our computation.  Accordingly, the temperature-dependent PL of a WSe$_{2}$-ML presents the higher rising temperature $T_1\sim 130$K but lacks the descending feature at or below the room temperature ($T_2\sim 645$K is predicted according to our simulation), which is consistent with the previous studies of refs \citenum{malic2018dark} and \citenum{selig2018dark} and seemingly better interprets another experiment of WSe$_{2}$-ML presented in ref~\citenum{arora2015excitonic}. 
The measured results of TMDC-MLs are known to vary sample by sample since the real samples of atomically thin TMDC-MLs are inevitably under and senstive to the various uncertain influences from the environment, e.g. the couplings from substrates or encapsulating materials \cite{ugeda2014giant,ajayi2017approaching,cadiz2017excitonic,zhang2015probing}, and the variation in the lattice constant \cite{hsu2017evidence}. Thus, the studies of the free-standing TMDC-MLs in this work is not set to establish the quantitative determination of the electronic and excitonic structures of TMDC-MLs, but to offer a guiding principle to establish the mapping between the $T$-characteristics of the measured temperature-dependent PL and the underlying exciton fine structures.


As the $T_2$-characterized descending feature of the temperature-dependent PL intensity is recognized as the signature of high lying DX states, it is not clear yet which types of the high DX states, likely the intra-valley MF DX states out of the light cone, the inter-valley $KK'$ or $KQ$ DX ones, are the main causes of the $T_2$-characteristic feature. Next, we shall introduce strain as a tuning parameter to tailor the energy order of the BX and various DX bands for further manifestation of the $T_2$-signature.

\subsection{Strained WSe$_2$-MLs}

\subsubsection{Strain-dependent quasi-particle band structures}

Figure~\ref{fgr:BandStructuresWithBiaxialStrain}a-c shows the calculated quasi-particle band structure of a WSe$_2$-ML under the symmetric bi-axial stress varied to yield the compressive, slightly compressive, and tensile strains, $\epsilon_b=-1\%$, $\epsilon_b=-0.4\%$ and $\epsilon_b=+1\%$. In the DFT computation, we first adjust the lattice constants of the stressed WSe$_2$-ML according to the specified strain and then allows the individual ions within the unit cell to relax to determine the optimized lattice structure under the stress. With the tensile strain, the $Q$ valleys of the conduction band are shown remaining at the higher energy than the conduction $K$ and $K'$ valleys, i.e. $\Delta\epsilon_{c_1,c_2}^{QK}>0$. Turning the strain to be compressive, the energy of the conduction $Q$ valleys are shifted down while that of the conduction $K$ valley is raised. As the strain is tuned to be $\epsilon_b=-0.4\%$,  the band edges of the conduction $Q$ and $K$ valleys are aligned up ($\Delta\epsilon_{c_1,c_2}^{QK} = 0$). Further increasing the compressive stress ($\epsilon_b < -0.4\%$), the conduction $Q$ valleys turn out to be energetically lower than the $K$ and $K'$ ones ($\Delta\epsilon_{c_1,c_2}^{QK}<0$). It is shown that the valley structure of the conduction band of a WSe$_2$-ML can be effectively tailored by applying a tuning symmetric bi-axial stress, featured by the opposite energy shifts of the $K$- and $Q$-valleys made by the stress \cite{hsu2017evidence,khatibi2018impact,aslan2018strain}. Thus, with the stress-tuned conduction valleys, the level energies of the intra-valley $KK$ and inter-valley $KQ$ exciton bands can be manipulated and even re-ordered by the applied stress.

\subsubsection{Strain-dependent exciton band structures}

Based on the strain-dependent quasi-particle band structures, the excitonic band dispersions of the WSe$_2$-MLs under the bi-axial stress are calculated. Figure~\ref{fgr:BandStructuresWithBiaxialStrain}d shows the band-edge energies of the intra-valley BX, the intra-valley SF DX, and the $KQ$ inter-valley MF DX as functions of the strain ranged between $\epsilon_b=\{ -1\%, +1\% \}$. \cite{frisenda2017biaxial}.
Because of the opposite trends of the energy shift of the conduction $K$- and $Q$-valley energies versus strain, the applied bi-axial stress affects the energies of the inter-valley and intra-valley excitons differently. The former increases with increasing the tensile stress, while the latter remains insensitive to the external stress. Thus, the energy of inter-valley $KQ$ exciton states can be varied from below (compressive strain), nearby (slightly compressive strain), and above (tensile strain) the BX states in the small range of strain, $\epsilon_b=\{ -1\%, +1\% \}$.  With a slightly compressive strain ($-0.3\% < \epsilon_{b} < -0.1\%$), the energy band of the inter-valley MF DX might lie between those of the intra-valley BX and SF DX.

\subsubsection{Strain-dependent PL spectra}

Figure~\ref{fgr:PLpeaksAndSignatures}a shows the intensities of the main PL peaks, as functions of $T$, of the stressed WSe$_2$-ML with $\epsilon_{b}=- 0.5\%$, $\epsilon_{b}=- 0.2\%$, $\epsilon_{b}= + 0.1\%$ and $\epsilon_{b}= + 0.4\%$, named by the cases A, B, C, and D, respectively.
The four representative cases possess the distinctive exciton fine structures where the band-edge energies of the intra-valley BX, and the intra-valley and inter-valley DXs are differently ordered.
In the case A (B), the energy of the inter-valley $KQ$ DX lies below (between) those of the SF DX and BX. In the case C (D), the energy of the inter-valley $KQ$ DX is slightly higher (much higher) than those of the both SF DX and BX.

In Figure~\ref{fgr:PLpeaksAndSignatures}a, one can observe how strain-dependent characteristics ($T_1$, $T_2$, and $\gamma_T$) evolves against varying the strain to undergo the situations of four representative cases.
First, one observes the high rising temperature $T_1 \sim 145$K in the compressive strain case with $\epsilon_{b}=- 0.5\%$.  As shown in the previous analysis, the rising temperature $T_1$ reflects the energy difference between the BX and the lowest DX levels. The high $T_1$ of the case A results from the large energy difference between the intra-valley BX and the lowest inter-valley MF DX whose energy is pulled down much by the compressive stress (See the regime I in Figure~\ref{fgr:BandStructuresWithBiaxialStrain}d). In the other three cases B, C, and D where $\epsilon_b > -0.2\%$, the rising temperatures $T_1\sim 65$K become lower and similar because in those cases the strain-insensitive intra-valley SF DX band remains as the lowest exciton band, with the similar energy difference from the high BX one (See the regimes II and III in Figure~\ref{fgr:BandStructuresWithBiaxialStrain}d).

Figure~\ref{fgr:PLpeaksAndSignatures}b shows the distinctive characteristics, $T_1$, $T_2$, and $\gamma_T$, extracted from the temperature-dependent PLs of the stressed WSe$_2$-ML versus the strain.
Compared with Figure~\ref{fgr:BandStructuresWithBiaxialStrain}d, one can find that the $T_1$ presented in the upper panel follows the similar strain-dependence of the energy difference between the BX and the lowest DX, which changes abruptly as the SF and MF DX levels cross at $\epsilon_{b}=- 0.3\%$.
Correspondingly, an abrupt change of the strain-dependent $T_1$ is also observed at $\epsilon_{b}=- 0.3\%$ in Figure~\ref{fgr:PLpeaksAndSignatures}b. From both analysis and numerical results, the rising temperature $T_1$ is firmly recognized as the signature about the lowest DX states and used to measure the energy difference between the band-edge energies of the BX and the lowest DX even in the presence of other high but nearby MF DX bands.

Next, we turn to discuss the high-temperature signature, $T_2$. In Figure~\ref{fgr:PLpeaksAndSignatures}a, the descending features in the high-temperature regime ($T\sim 300$K) are observed in the cases B, C, and D, but not in the case A. From the previous model analysis, the descending feature should be associated with some high lying DX states with multi-fold degeneracy. The missing of the $T_2$ feature in the case A where the inter-valley $KQ$ DX states are pushed down below the BX ones by the applied compressive stress confirms the scenario.

The middle panel of Figure~\ref{fgr:PLpeaksAndSignatures}b shows the $T_2$ extracted from the temperature-dependent PL spectra of the stressed WSe$_2$-ML, as a function of the strain.
One notes that, only as $\epsilon_{b} \gtrsim - 0.1\%$ where the inter-valley $KQ$ DX bands are not below the BX states (e.g. the cases C and D), the $T_2$ falls into the observable room-temperature regime. With $\epsilon_{b} < - 0.3\%$, where the both of the intra-valley SF DX and the inter-valley MF DX are in energy lower than the intra-valley BX (e.g. the case A), the $T_2$ characteristic is absent. With $- 0.3\% < \epsilon_{b} < - 0.1\%$, where the inter-valley $KQ$ DX bands lie between the BX and SF DX states, i.e. the case B,  the $T_2$ characteristic emerges but is overall still high.

To further confirm the dominant role of the MF $KQ$ DX bands in the descending PL intensity, we re-calculate the temperature-dependent PL spectra of the WSe$_2$-ML but without the inclusion of the band segments of the $KQ$ inter-valley MF DX.
For the identification of the $T_{2}$-effects from the various high-lying DXs, we carry out the test computation where the intra-valley SA and MF DX states resident out of the light cone, $KK'$ inter-valley MF DXs, (and the $KQ$ inter-valley MF DXs as well,) are artificially removed and compare the yielded $T_2$, as represented by the gray (black) line in the middle panel of Figure~\ref{fgr:PLpeaksAndSignatures}b, with that (the blue line) from the full computation.
One can see that, even with the removal of the all MF DX states but preserving the $KQ$ inter-valley MF DXs, the $T_2$ given from the test computation presents the similar non-linear strain-dependence as that from the full computation, which can reach or even below the room temperature with tensile or no strain.
By contrast, in the absence of all MF DX states including the $KQ$ inter-valley  MF DX ones, the $T_2$ turns out to decrease monotonically with increasing the strain but remain always higher than the room-temperature.  This confirms that the turning temperature $T_2$ is truly the observable signature in the room-temperature regime about the high lying MF DX bands, distinctive from $T_1$ relevant only to the lowest SF DX. By the comparison of the behaviors of the $T_2$ from the two test computations, the two types of MF DXs, intra-valley SA and MF DX and the $KK'$ inter-valley MF one, are realized to contribute the $T_2$-characteristic descending feature as well but not so much as the inter-valley $KQ$ DXs.

In summary, we present a theoretical and computational investigation of the excitonic band structures and the temperature-dependent time-integrated photo-luminescence of WSe$_2$-MLs by solving the BSE with the consideration of both {\it e-h} direct and exchange Coulomb interactions in the tight-binding scheme established from the first principle density-functional-theory computation. As compared with other TMDC-MLs, the band structure of WSe$_2$-ML exhibits even more pronounced valley characteristics, featured by not only the common $K$ and $K'$ valleys, but also the six-fold degenerate $Q$ conduction ones. This leads to the rich spin- and valley-related fine structures of the low-lying exciton bands of a photo-excited WSe$_2$-ML, where three main types of excitons, i.e. the intra-valley BX, intra-valley SF DX (the lowest exciton states) and various inter-valley ($KK'$, $KQ$, ...) MF DXs, are involved. The latter two are optical invisible but their emergences in the low-energy regime of the excitonic spectrum of a WSe$_2$-ML are yet essential in the temperature dependent photo-luminescence. To elucidate the underlying physics, a three-level model is employed to analyze the thermal statistics in the exciton states, under the influences from both of the low-lying SF DX and the high-lying multi-fold degenerate MF DX states, which advices us in the search for the signatures of the optically invisible SF and MF exciton states in the temperature-dependent photo-luminescences. With the supportive analysis, we reveal, in addition to the commonly known low-temperature signature of the lowest SF DX, the other high-temperature signature arising from the high lying MF DX states in the temperature-dependent photo-luminescence of a WSe$_2$-ML. The signature of the MF DXs is characterized by a turning temperature $T_2 \sim 263$K where the photo-luminescence intensity turns to descend slowly with increasing the temperature, while the lowest SF DX states is evidenced by the low-temperature signature, $T_1 \sim 66$K, at which the PL intensity turns out to rise up rapidly. The model analysis derives explicitly the condition for the emergence of the $T_2$ signature, which requires the high degeneracy of the high lying dark exciton bands, fulfilled exactly by the twelve-fold degenerate inter-valley MF $KQ$ DX bands. At last, by introducing strain in the stress-controlled WSe$_2$-ML as a tuning parameter to tailor the energy order of the $K$ and $Q$ conduction valleys, the $T_1$- and $T_2$- characteristics are manifested as the true signatures of the SF and MF DXs, respectively, and their emergence and identification allow one to infer the energy locations of the DXs with respect to that of the BX.

\begin{acknowledgement}


This study is supported by the Ministry of Science and Technology, Taiwan, under Research of Excellence (RoE) Program (MOST-107-2633-E-009-003 and MOST 108-2633-E-009-001) and the contracts (MOST-106-2112-M-009-015-MY3, MOST-106-2221-E-009-113-MY3), and by National Center for High-Performance Computing (NCHC), Taiwan.

\end{acknowledgement}

\begin{suppinfo}


The Supporting Information is available free of charge on the
ACS Publications website at DOI:xxxxxxx

The technical information about the computation of the quasi-particle band structures using the DFT \textbf{\footnotesize{VASP}} package, the wannierization of electron Bloch wave functions using the \textbf{\footnotesize{WANNIER90}} package, the formalism of momentum matrix elements, the numerical implementation of the Bethe-Salpeter equation,
the numerical method for the calculation of the {\it e-h} direct and exchange Coulomb interactions
, the model analysis for the thermal population of exciton states,
the statistics of the reported conduction band splitting of WX$_2$-ML in literature, and supporting figures.

\end{suppinfo}


\bibliography{NanoLett_Ref}

\begin{figure}
\centering
\hbox{\hspace{-2cm} \includegraphics[width=1.2\textwidth]{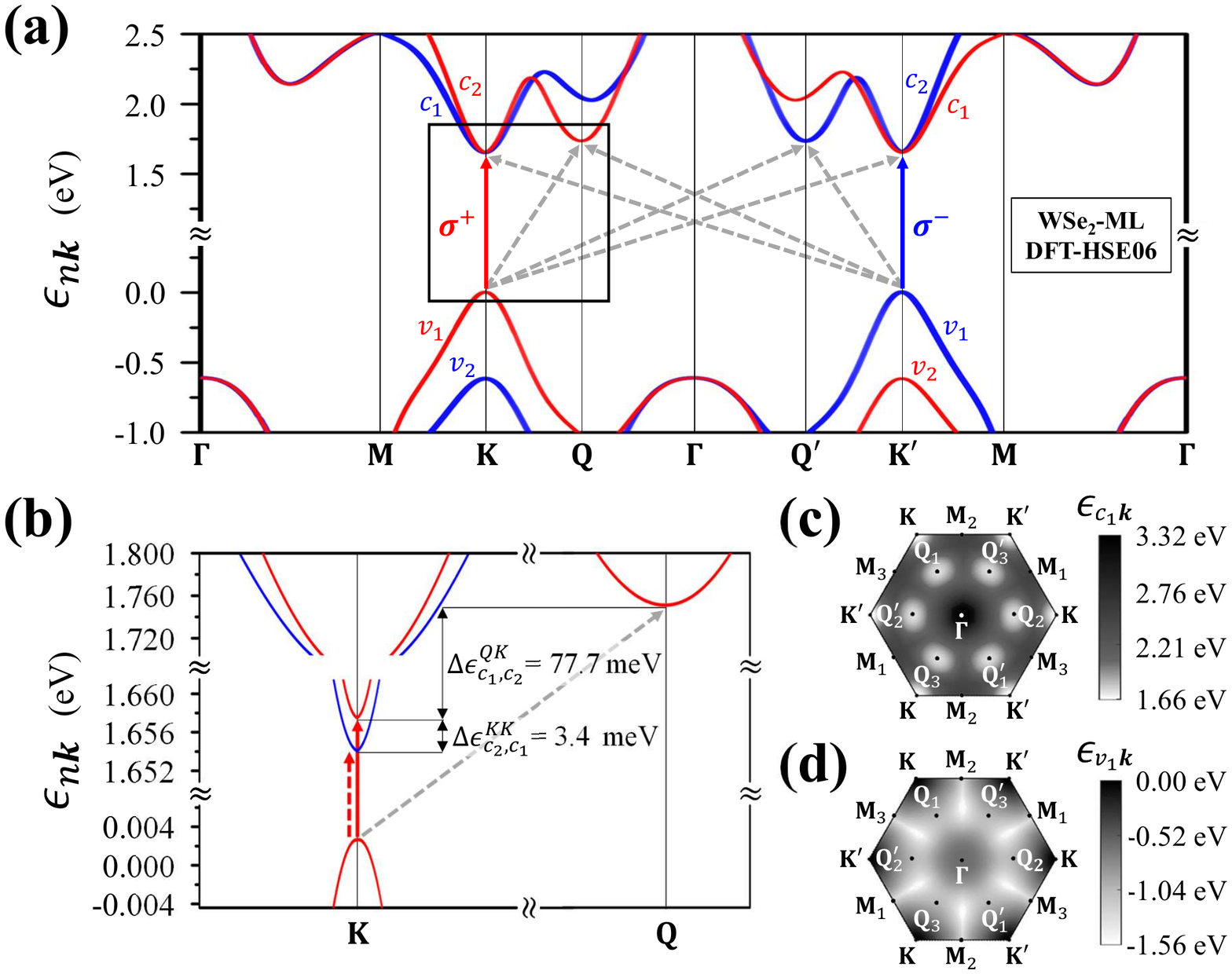}}
\caption{The quasi-particle band structure of WSe$_2$-ML calculated by using the HSE06 exchange-correlation functionals in the DFT. (a) The band structure along the line connecting the points, $\Gamma$, $M$, $K$, and $Q$, in the reciprocal space. The red (blue) bands indicate the spin-up (spin-down) states, the red (blue) arrow line indicates the direct right-circularly polarized $\sigma^{+}$- (left-circularly polarized $\sigma^{-}$-)transition, and the gray arrow lines indicate the indirect momentum-forbidden (MF) transitions. (b) The enlarged view of the quasi-particle band structures near the $K$- and $Q$-valleys. The red dashed and solid arrow lines indicate the spin-forbidden (SF) and spin-allowed (SA) transition, respectively. (c) and (d) indicate the energy contour of the lowest conduction band $c_{1}$ and the highest valence band $v_{1}$ inside the 1st Brillouin zone, respectively.}
\label{fgr:QuasiParticleBandStructure}
\end{figure}

\begin{figure}
\centering
\includegraphics[width=1.0\textwidth]{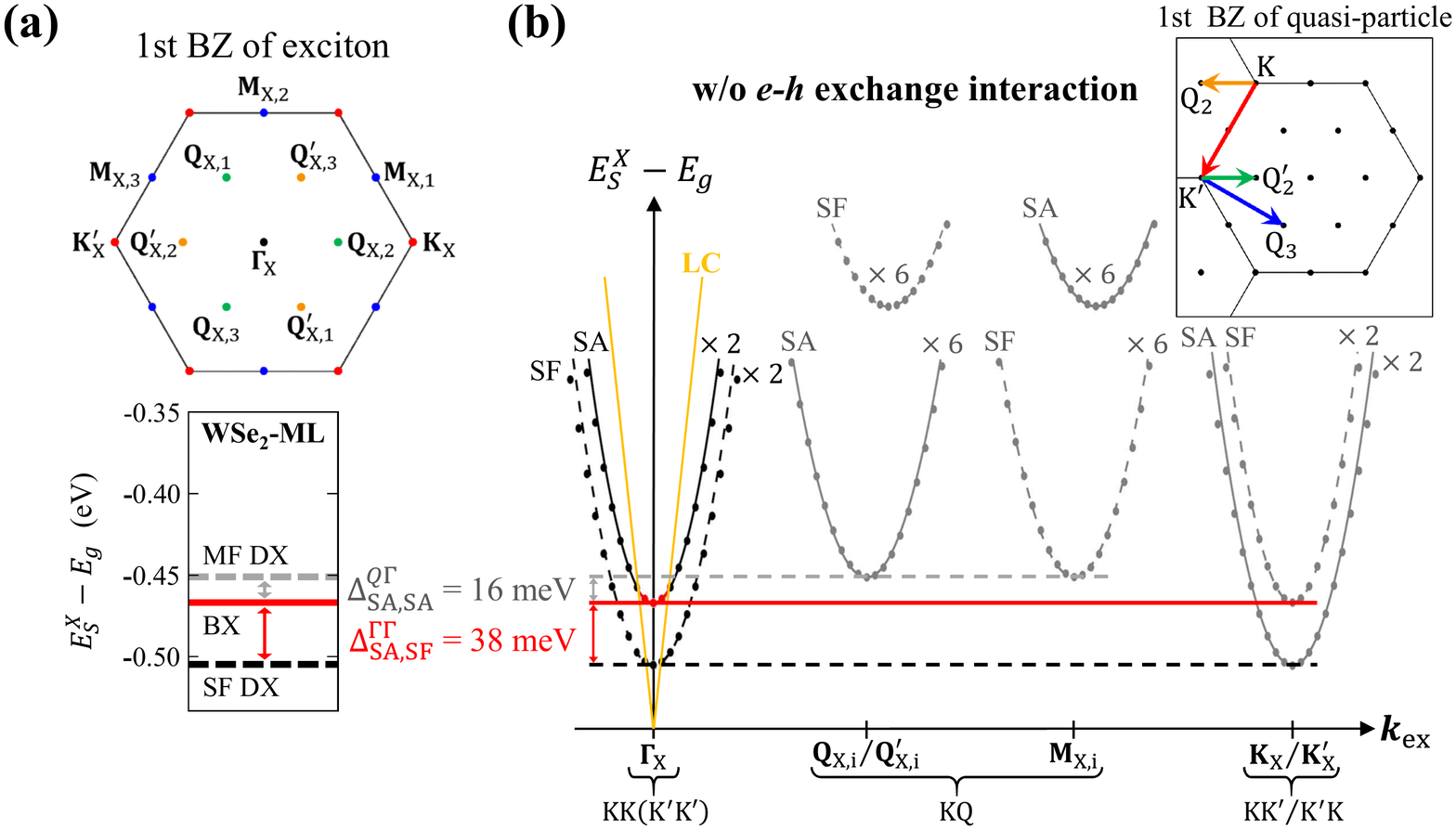}
\caption{The exciton band structure of a WSe$_2$-ML calculated by solving the BSE based on the quasi-particle band structure of Figure~\ref{fgr:QuasiParticleBandStructure}
and with the neglect of the {\it e-h} exchange interaction.
(a) Top: the first Brillouin zone (1st BZ) in the $\VEC{k}_{ex}$ reciprocal space. Bottom: the level energies corresponding to the band-edges of the intra-valley SF DX (black dashed line) and the intra-valley SA BX (red solid line) at the $\Gamma_X$ or $K_X/K_X'$ exciton valleys, and the inter-valley MF DX (gray dashed line) at the $Q_{X}/Q_{X}'$ and $M_X$ valleys presented in (b). (b) The calculated exciton band dispersions around the valleys, $\Gamma_X$, $Q_{X}/Q_{X}'$, $M_X$, and $K_X/K_X'$, in the $\VEC{k}_{ex}$ reciprocal space of exciton. The black (gray) curves describe the band dispersions of the intra(inter)-valley exciton states, and the solid (dashed) lines indicate the spin-allowed (spin-forbidden) exciton states.  The bright exciton states inside the light-cone (LC) that are spin- and momentum-allowed for the optical transitions are highlighted in red color. Note that the light cone (LC) dispersion in orange color is depicted schematically. The degeneracy number are marked next to each exciton bands. Inset: the vectors corresponding to the special points in the 1st BZ of exciton are depicted in the corresponding colors in the BZ of quasi-particle. }
\label{fgr:ExcitonBandStructure}
\end{figure}

\begin{figure}
\centering
\includegraphics[width=1.0\textwidth]{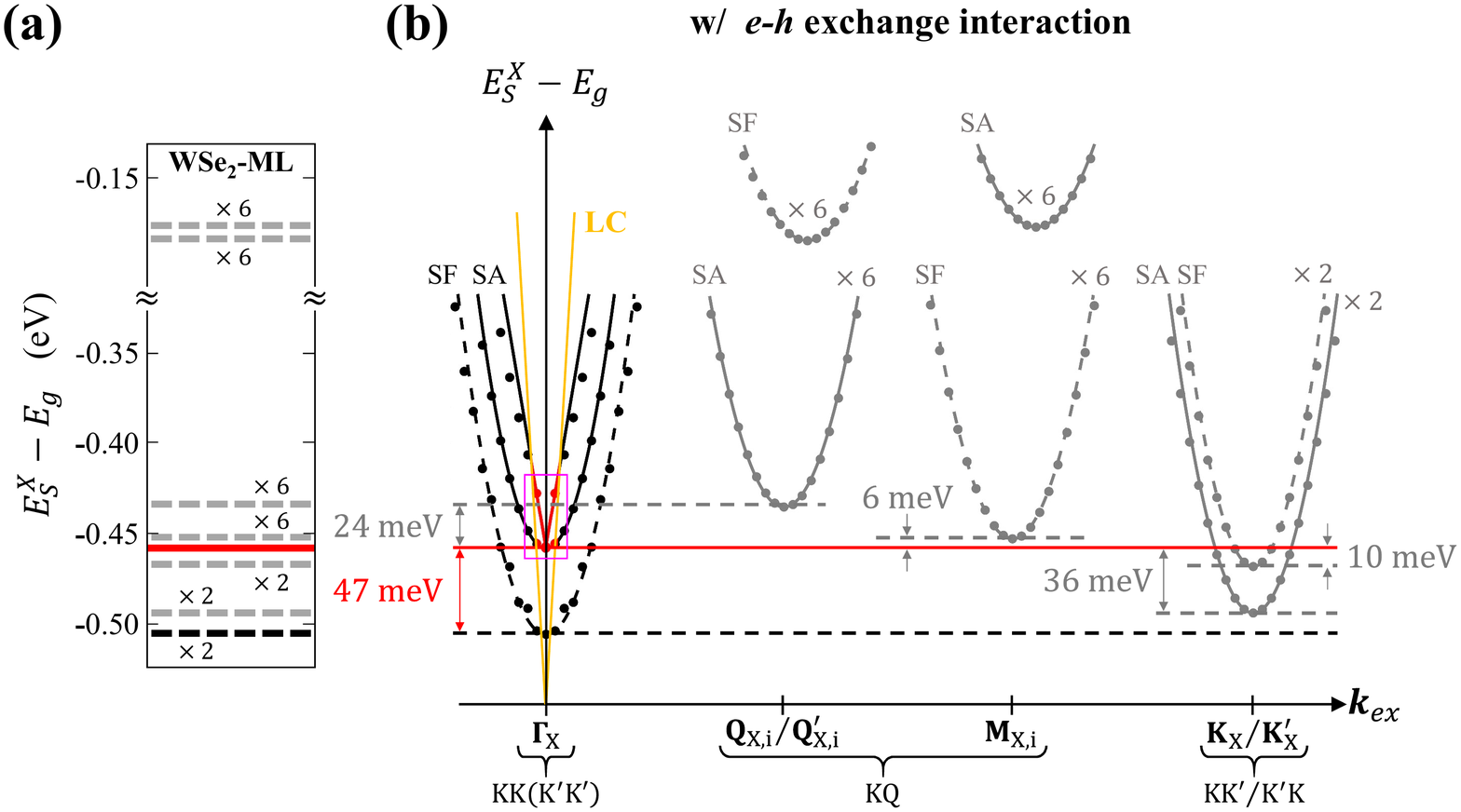}
\caption{
Same as Figure~\ref{fgr:ExcitonBandStructure}, but with the additional consideration of the {\it e-h} exchange interaction. One notes that the {\it e-h} exchange interaction gives rise to the energy splitting of the bright exciton band, further separates the levels of the bright exciton and the lowest spin-forbidden dark exciton bands, and lifts the degeneracies of the inter-valley $KQ$ and $KK'/K'K$ dark exciton bands. The zoomed-in view of the spin-split bright exciton bands specified by the magenta box is shown in Figure~S4 of the SI.
}
\label{fgr:ExchangeExcitonBandStructure}
\end{figure}

\begin{figure}
\centering
\includegraphics[width=1.0\textwidth]{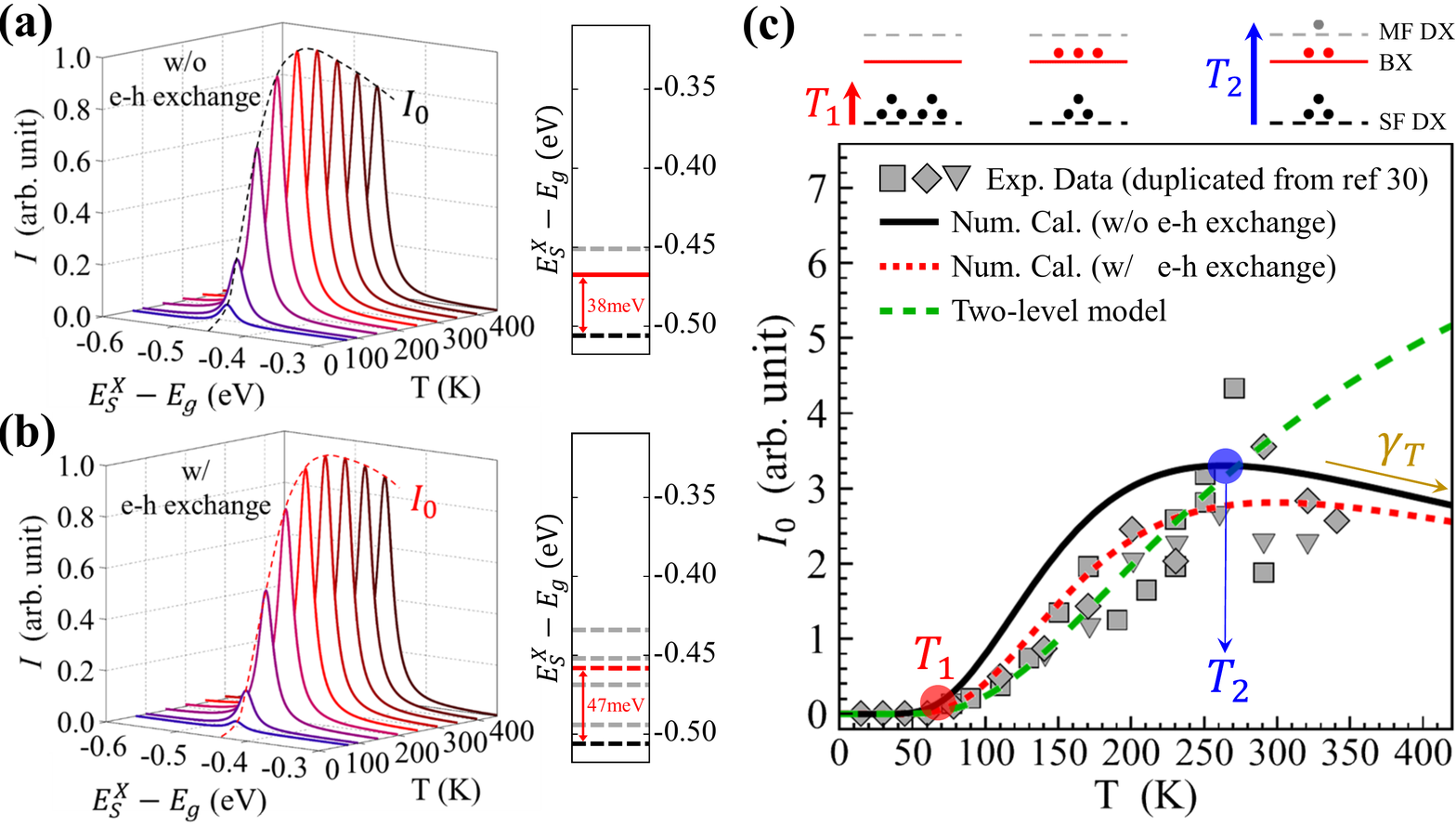}
\caption{(a) The numerically calculated PL spectrum of a WSe$_2$-ML under thermalization at the temperatures $T=75, 100, 150, ..., 400$K,
based on the BSE where the {\it e-h} exchange interaction is not included. Inset: The energy levels of the low-lying spin- and valley-resolved exciton band edges.  (b) Same as (a), but with the consideration of the {\it e-h} exchange interaction in the BSE. The inset shows that the {\it e-h} exchange interaction gives rise to more fine structures of the exciton spectrum, and makes the increase of the energy difference between the intra-valley bright- (BX) and spin-forbidden dark exciton (SF-DX) band edges from 38meV to 47meV.
(c) The temperature-dependent intensity of the main PL peak ($I_{0}$) extracted from (a) [(b)], represented by the black solid line [red dashed line]. The temperature dependence of $I_{0}$ is characterized by the rising temperature $T_1$ (the turning temperature $T_2$) being the signature of the lowest SF-DX states (the high lying MF-DX states), where the temperature-dependent PL intensity turns to rise up (descend). $\gamma_{T}$ stands for the descending rate of $I_{0}$ with respect to $T$.
The green dashed line shows the thermal population of BX simulated by the widely used two-level model (See Figure~\ref{fgr:ThreeLevelModel}a). As discussed in the main text, the simplified two-level model can well fit the low-temperature behavior of the temperature-dependent PL intensity but cannot describe the high-temperature descending feature characterized by $T_2$. Corresponding to the low-, moderate-, and high-temperature regimes, the schematics of the thermal populations on the BX and DX states are depicted above the plot.
The experimental data of ref~\citenum{zhang2015experimental} measured for three samples (square, diamond, and triangle symbol in gray color) are duplicated here for comparison with the theoretical simulations.
}
\label{fgr:TemperatureDependentPLIntensity}
\end{figure}

\begin{figure}
\centering
\includegraphics[width=1.0\textwidth]{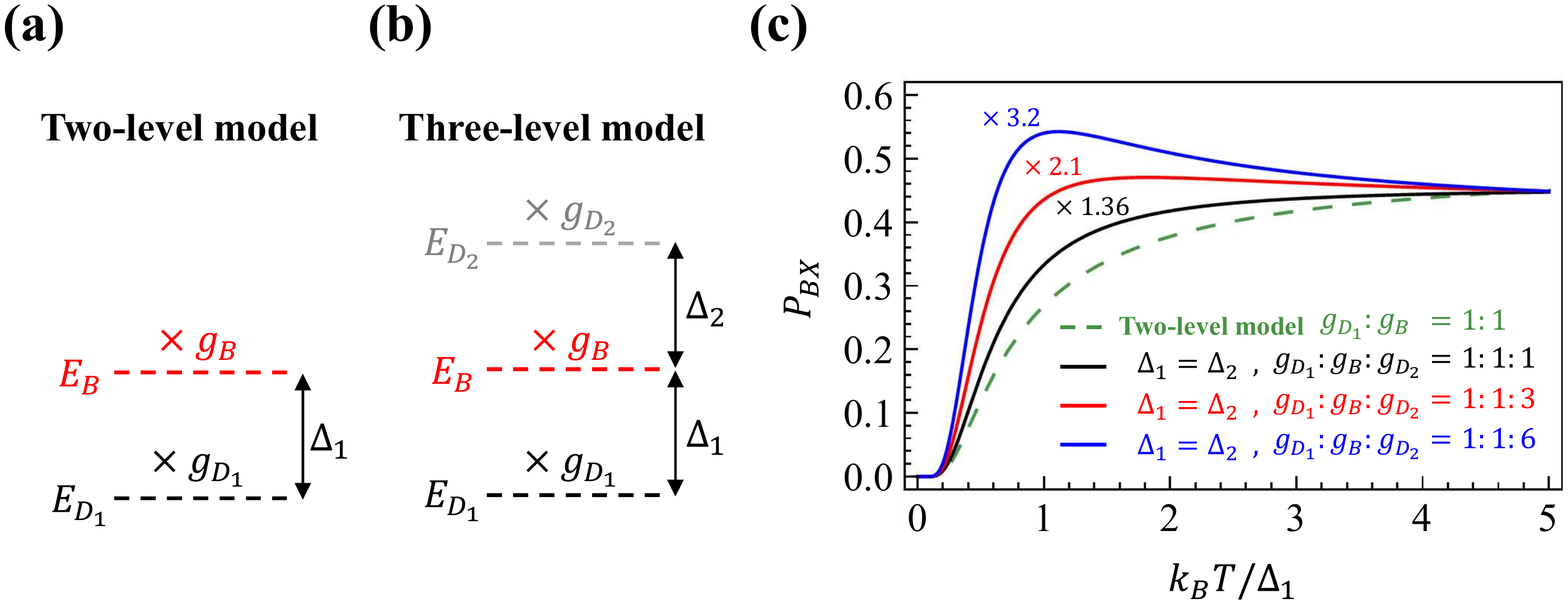}
\caption{The schematics of the exciton models used for the analysis of the thermal population of exciton states and the temperature dependent PL intensity. (a) The two-level model composed of a low DX level ($D_1$) and a high BX level ($B$) with the energy difference $\Delta_1=E_B-E_{D_1} >0$. (b) The three-level model that is extended from the two-level one with one additional DX level ($D_2$) at the energy differing from the BX one by $\Delta_2=E_{D_2}-E_{B}$. The degeneracies of the three exciton levels are denoted by $g_{D_1}$, $g_{B}$, and $g_{D_2}$, respectively. (c) The simulated thermal population of the BX level  in the Boltzmann statistics calculated by using the two- and three-level models, where $\Delta_1=\Delta_2$ and various ratios of the level degeneracies (as indicated in the panel) are taken. From the model simulation, the descending feature at high-temperature is shown to be simulated by the three-level model but not by the two-level one, and is more pronounced as the degeneracy of the high DX level ($g_{D_2}$) is greater than those of the other low-lying levels. }
\label{fgr:ThreeLevelModel}
\end{figure}

\begin{figure}
\centering
\includegraphics[width=1.0\textwidth]{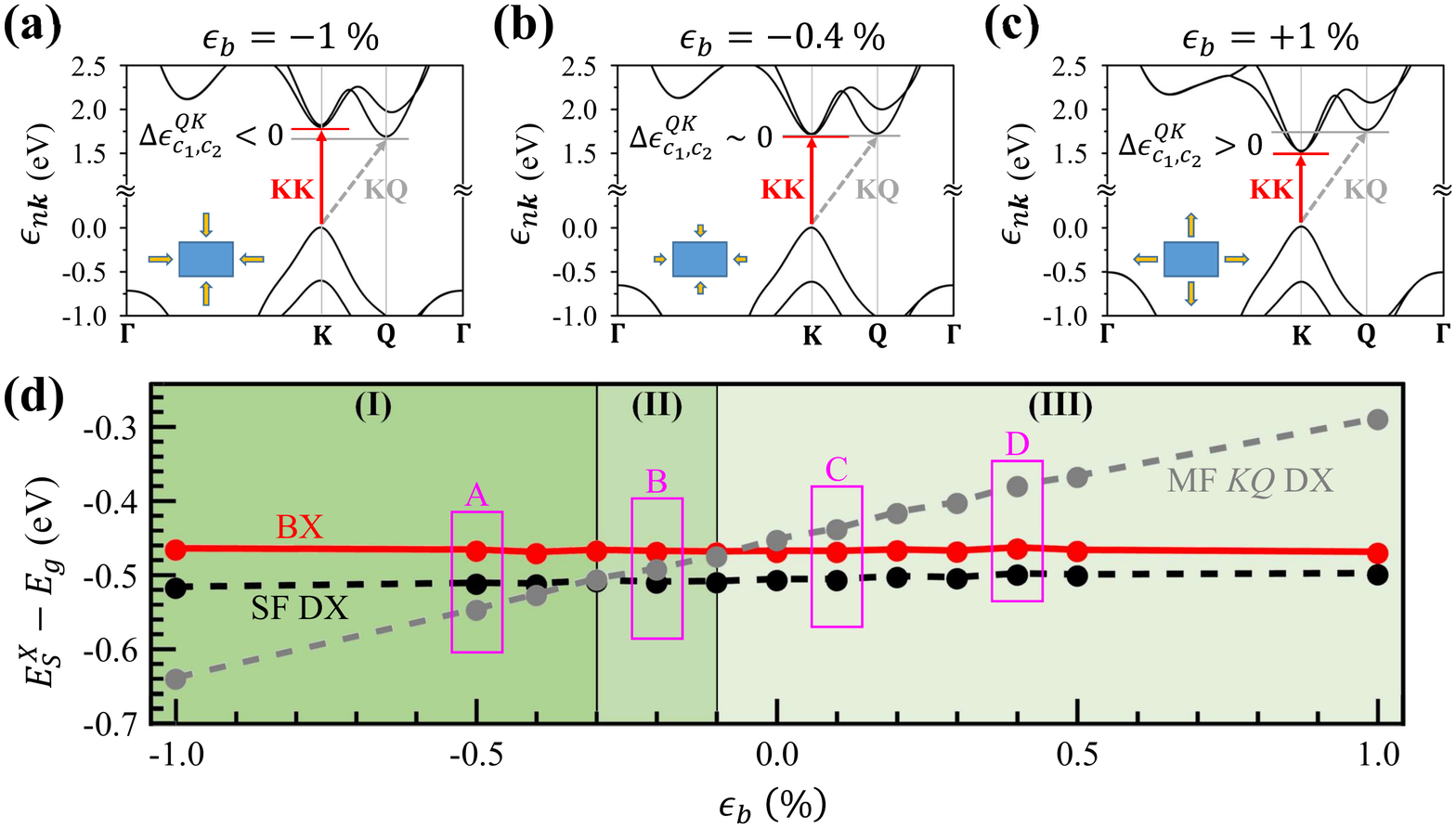}
\caption{The quasi-particle band structures of strained WSe$_2$-MLs with (a) the compressive ($\epsilon_b=-1\%$), (b) the slightly compressive ($\epsilon_b=-0.4\%$), and (c) the tensile ($\epsilon_b=+1\%$) symmetric bi-axial strain.
The energy difference between the direct (red arrow lines) and indirect (gray dashed arrow lines) non-interacting particle transitions, given by $\Delta \epsilon_{c_{1},c_{2}}^{QK} \equiv \epsilon_{c_{1} \VEC{Q}}-\epsilon_{c_{2}\VEC{K}}$, is positive (negative) as the WSe$_2$-ML is subjected to a tensile (compressive) strain. In between, the energies of the direct and indirect transitions are nearly equal, i.e. $\Delta \epsilon_{c_{1},c_{2}}^{QK} \sim 0$, with the  slightly compressive strain, $\epsilon_b=-0.4\%$. (d) The band-edge energies of the intra-valley SA BX (red), the intra-valley SF DX (black), and the $KQ$ inter-valley MF DX (gray) of the stressed WSe$_2$-ML as a function of the bi-axial strain.  According to the band-edged energy order of the three types of the exciton, three strain regimes named by (I), (II) and (III) are defined. In the regime (I), the inter-valley MF DX bands are lowered by the compressive strain to be the lowest exciton states of the stressed WSe$_2$-MLs. In the regime (II), with slightly compressive strain ($-0.3\%<\epsilon_b<-0.1\%$) the level energy of the inter-valley MF DX bands lie between those of the intra-valley BX and SF DX bands. In the regime (III), the inter-valley MF DX bands remain higher than the both of the intra-valley BX and SF DX ones. Four specific cases labeled by A, B, C, and D and highlighted by magenta boxes are selected as the case studies presented in Figure~\ref{fgr:PLpeaksAndSignatures}a.}
\label{fgr:BandStructuresWithBiaxialStrain}
\end{figure}

\begin{figure}
\centering
\includegraphics[width=1.0\textwidth]{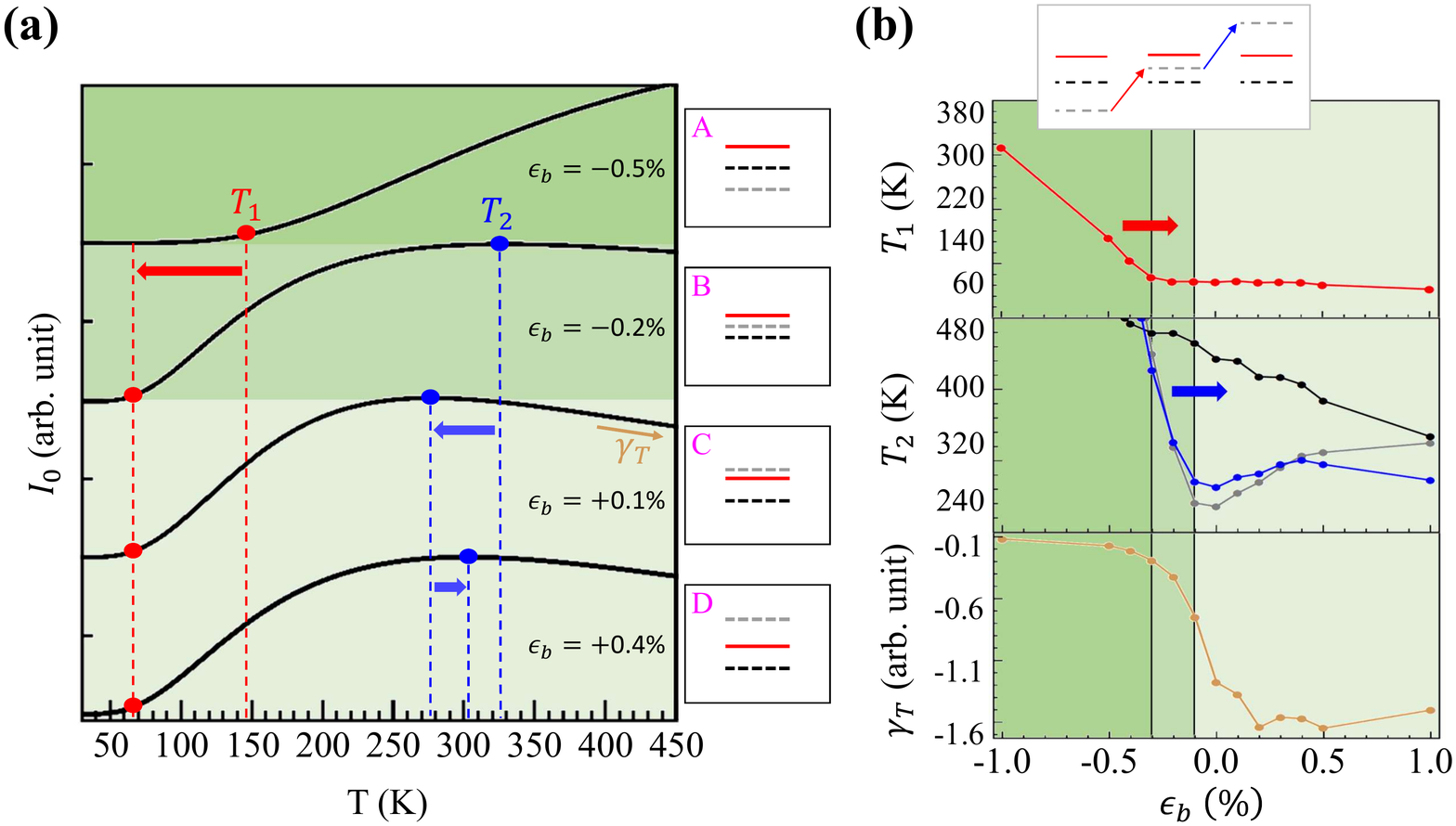}
\caption{(a) The intensities of the main PL peaks of the strained WSe$_2$-MLs as functions of the temperature, with the strain $\epsilon_{b}=-0.5\%$ (case A), $\epsilon_{b}=-0.2\%$ (case B), $\epsilon_{b}=+0.1\%$ (case C), and $\epsilon_{b}=+0.4\%$ (case D). The four cases are indicated by the magenta boxes in Figure~\ref{fgr:BandStructuresWithBiaxialStrain}d as the representative cases of the distinctive relative energy locations of the inter-valley MF DX with respect to those of the intra-valley BX and SF DX, and exhibit differently the characteristic $T_1$-, $T_2$-temperatures, and descending rates $\gamma_T$ of the temperature-dependent PL intensities. (b) The strain-dependent $T_1$, $T_2$, and $\gamma_T$ extracted from the temperature-dependent PLs of the WSe$_2$-ML with the varying bi-axial strain, $\epsilon_b=\{-1\%,+1\%\}$. In the middle panel, the $T_{2}$ obtained from the full computation is represented by the blue line. For the identification of the $T_{2}$-effects from the various high-lying DXs, the $T_{2}$ from the test computation where the intra-valley SA exciton states resident out of the light cone, $KK'$ inter-valley DXs, (and the $KQ$ inter-valley DXs,) are removed is represented by the gray (black) line for comparison.}
\label{fgr:PLpeaksAndSignatures}
\end{figure}

\newpage

\begin{table}
  \caption{The effective masses of the quasi-particle band extrema (in units of free electron mass).}
  \label{tbl:ElectronicMass}
  {
  \begin{tabular}{|c|c|c|c|c|}
    \hline
    $m_{v_{1},K (K')}$  & $m_{c_{1},K (K')}$ & $m_{c_{2},K (K')}$ & $m_{c_{1},Q (Q')}$ & $m_{c_{2},Q (Q')}$  \\
    \hline
    -0.33 & 0.37 & 0.25 & 0.53 & 0.95  \\
    \hline
  \end{tabular}}
\end{table}

\begin{table}
  \caption{The effective masses of the excitonic band extrema (in units of free electron mass).}
  \label{tbl:ExcitonicMass}
  \begin{tabular}{|c|c|c|c|c|c|c|c|}
    \hline
    $M_{SF,\Gamma_{X}}^{X}$ & $M_{SA,\Gamma_{X}}^{X}$ & $M_{SA,Q_{X,i}(Q'_{X,i})}^{X}$ & $M_{SF,Q_{X,i}(Q'_{X,i})}^{X}$ & $M_{SF,M_{X,i}}^{X}$ & $M_{SA,M_{X,i}}^{X}$ & $M_{SA,K_{X}(K'_{X})}^{X}$ & $M_{SF,K_{X}(K'_{X})}^{X}$ \\
    \hline
    0.878 & 0.714 & 1.027 & 1.618 & 1.030 & 1.614 & 0.883 & 0.716  \\
    \hline
  \end{tabular}
\end{table}

\end{document}